\documentclass[a4paper,fleqn,usenatbib]{mnras}

\usepackage{mathptmx}

\usepackage[T1]{fontenc}
\usepackage{ae,aecompl}

\usepackage{graphicx}	
\usepackage{amsmath}	
\usepackage{amssymb}	
\usepackage{threeparttable}
\usepackage{verbatim}
\usepackage{listings}
\newcommand{\teff}{\ensuremath{T_{\mathrm{eff}}}}
\newcommand{\logg}{\ensuremath{\log g}}
\newcommand{\feh}{\ensuremath{[\rm Fe/\rm H]}}

\newcommand\nion[2]{#1\,\lowercase{{\sc #2}}}
\newcommand\nionb[2]{#1\,\lowercase{{\small #2}}}

\newcommand{\kms}{km s$^{-1}$}
\newcommand{\Vmic}{v_{\rm{mic}}}

\newcommand{\swoc}{{\tt SWOC}}

\title[Optimizing spectral wavelength coverage: SWOC]{A new algorithm for optimizing the wavelength coverage for spectroscopic studies: Spectral Wavelength Optimization Code (SWOC)}
\author[Ruchti et al.]{G.~R.~Ruchti,$^{1}$\thanks{{\tt Email: greg@astro.lu.se}}
S. Feltzing,$^{1}$
K. Lind,$^{2, 3}$
E. Caffau,$^{4}$
A. J. Korn,$^{3}$
O. Schnurr$^{5}$
\newauthor
C. J. Hansen,$^{6}$
A. Koch,$^{7}$
L. Sbordone,$^{8,9}$
R. S. de Jong$^{5}$
\\
$^{1}$Lund Observatory, Department of Astronomy and Theoretical Physics, Box 43, SE-22100 Lund, Sweden\\
$^{2}$Max-Planck Institut f\"{u}r Astronomie, K\"{o}nigstuhl 17, 69117 Heidelberg, Germany\\
$^{3}$Department of Physics and Astronomy, Uppsala University, Box 516, SE-75120 Uppsala, Sweden\\
$^{4}$GEPI, Observatoire de Paris, CNRS, Universit\'e Paris Diderot, 5 Place Jules Janssen, 92190 Meudon, France\\
$^{5}$Astrophysikalisches Institut Potsdam, An der Sternwarte 16, D-14482 Potsdam, Germany\\
$^{6}$Landessternwarte, Zentrum f\"ur Astronomie der Universit\"at Heidelberg, K\"onigstuhl 12, D-69117 Heidelberg, Germany\\
$^{7}$Physics Department, Lancaster University, Lancaster LA1 4YB, UK\\
$^{8}$Millennium Institute of Astrophysics, Av. Vicu\~{n}a Mackenna 4860, 782-0436 Macul, Santiago, Chile\\
$^{9}$Pontificia Universidad Cat\'{o}lica de Chile, Av. Vicu\~{n}a Mackenna 4860, 782-0436 Macul, Santiago, Chile
}

\date{Accepted 2016 June 1. Received 2016 June 1; in original form 2016 May 10}

\pubyear{2016}

\begin{document}
\label{firstpage}
\pagerange{\pageref{firstpage}--\pageref{lastpage}}
\maketitle

\begin{abstract}

The past decade and a half has seen the design and execution of several ground-based spectroscopic surveys, both Galactic and Extra-galactic.  Additionally, new surveys are being designed that extend the boundaries of current surveys.  In this context, many important considerations must be done when designing a spectrograph for the future.  Among these is the determination of the optimum wavelength coverage.  In this work, we present a new code for determining the wavelength ranges that provide the optimal amount of information to achieve the required science goals for a given survey.  In its first mode, it utilizes a user-defined list of spectral features to compute a figure-of-merit for different spectral configurations.  The second mode utilizes a set of flux-calibrated spectra, determining the spectral regions that show the largest differences among the spectra.  Our algorithm is easily adaptable for any set of science requirements and any spectrograph design.  We apply the algorithm to several examples, including 4MOST, showing the method yields important design constraints to the wavelength regions. 

\end{abstract}



%
\section{Introduction}

Observational astronomy has entered an era of large-scale surveys, probing ever deeper into the Milky Way and the Universe.  Several imaging surveys are already ongoing or completed, such as SDSS \citep{eisenstein2011}, 2MASS \citep{skrutskie2006}, {\it WISE} \citep{wright2010}, and Skymapper \citep{keller2007}, with future surveys soon starting \citep[e.g. LSST;][]{ivezic2008}.  Further, at the end of 2013, the astronomical community celebrated the European Space Agency's launch of the {\it Gaia} astrometric space mission.  {\it Gaia} will deliver unparalleled high-precision positions and transverse velocities for a billion stars in the Milky Way \citep[][]{perryman2001,lindegren2008,lindegren2010}.  The imaging surveys and {\it Gaia} have led to the design and execution of ground-based, large-scale spectroscopic surveys (e.g. RAVE;  \citealt{steinmetz2006}, SEGUE; \citealt{yanny2009}, LAMOST; \citealt{cui2012}, GALAH/HERMES; \citealt{barden2010}, APOGEE; \citealt{eisenstein2011}, the {\it Gaia}-ESO Survey; \citealt{gilmore2012}).  The spectroscopic surveys deliver complementary data that, when combined with the astrometric data from {\it Gaia}, allow us to comprehensively map all of the major stellar components of the Milky Way.

Future European large-scale surveys, such as the 4-meter Multi-Object Spectroscopic Telescope (4MOST; \citealt{dejong2014}), WEAVE \citep{balcells2010}, and surveys conducted with the third generation VLT instrument MOONS \citep{cirasuolo2011} are also being planned, and are designed to provide the desired ground-based spectroscopic follow-up for {\it Gaia}.  These new data will provide an extraordinary resource and will enable transformative science by providing many of the critical observational inputs needed to dissect the formation history of the Milky Way and other galaxies in great detail.  However, these new surveys will come with limitations (e.g. in wavelength coverage and resolution) mainly due to constraints needed to be able to observe large numbers of targets.  Thus, it is important to perform a detailed investigation of the required performance of future spectrographs and also test the power of existing instruments.  Similar considerations can also be applied in other science areas where a spectrograph must be optimized.

There are many considerations that must be taken into account when designing a spectrograph for a given spectroscopic survey, e.g., design limitations such as the spectral resolution and sampling, the wavelength coverage, variation in fiber throughput for different wavelength regions, as well as the location in wavelength of the flux maximum for the typical target that will be observed by a given survey.  In this paper, we concentrate on determining the spectral coverage (for a given range in resolution) best suited for a given science case within a spectroscopic survey.  

Descriptions and arguments for how to determine the optimal placement of the wavelength bands are quite limited in the literature.  A recent example is \citet{caffau2013}, who discussed the optimal wavelength coverage, resolution, and sampling of an earlier 2-arm design for 4MOST.  More recently, \citet{hansen2015} performed a similar investigation for the preliminary design of the blue wavelength region ($\lambda < 4500$~\AA) of the high-resolution spectrograph on 4MOST.

In this paper, we present a simple and effective algorithm ({\tt SWOC}), which utilizes information contained within the spectra of different objects and/or the science requirements for a given survey to weigh among different combinations of short wavelength bands.  This program was originally conceived to optimize the wavelength coverage of the 4MOST spectrograph. However, it was written so that any spectrograph design can be evaluated.  It can also be used to decide on which settings to use for an existing instrument for a specific project.  Thus, the program can be applied to any spectroscopic study, both stellar and extra-galactic, on any spectrograph, at any resolution.

The paper is organized as follows.  In \S\ref{sec-swoc}, we present an overview of the program.  In \S\ref{sec-mode}, we describe the two different modes that can be used to evaluate the wavelength coverage.  We then present in \S\ref{sec-mw} several example analyses for the {\tt feature} mode, while and additional example of the {\tt spectral} mode is given in \S\ref{sec-eg}.  Finally, in \S\ref{sec-con}, we present our conclusions.

\section{The SWOC Program}
\label{sec-swoc}

The main goal of our methodology is to compute a figure-of-merit (FoM) for every possible combination of wavelength bands, which is based on the science goals intended to be achieved with a given spectrograph.  This FoM can then be used to:
\begin{itemize}
\item determine the optimal combination of wavelength bands, which give the maximum possible information from the spectrum.
\item assess the amount of information lost if a different combination of bands is used.
\item assess the ability of an existing spectrograph or spectrograph design to achieve the desired science goals.
\end{itemize}
To this end, we have developed a new Spectral Wavelength Optimization Code ({\tt SWOC}), which evaluates all possible combinations of wavelength bands to determine those that maximize the FoM.

\swoc{} was written as a package for the statistical software environment {\tt R}\footnote{\url{https://www.r-project.org}}.  {\tt R} is available as free software, thus allowing anyone to use \swoc{} without any software licenses.  The \swoc{} package can be obtained through a website\footnote{\url{http://www.astro.lu.se/~greg/swoc.html}} or upon request.
 
\section{The two modes of SWOC}\label{sec-mode}

\swoc{} operates in two separate modes, {\tt feature} and {\tt spectral} modes.  Input details for each mode are described in Appendix~\ref{sec-inswoc},  while common input definitions are listed in Table~\ref{tab-inputs}.  Additional help files can also be accessed within the package in {\tt R}.    

Here, we describe the methodology behind each mode. 
  
\subsection{Feature Mode}\label{sec-feat}

The main philosophy behind this mode is to make use of specific spectral features defined by the user to assess combinations of wavelength bands with a given scientific problem in mind.  The spectral features can, e.g., be atomic or molecular absorption/emission lines.  In order to find the best set of wavelength bands that optimizes the science, we define a figure-of-merit (FoM) based on the number of spectral features present within each set of wavelength configurations.   Different spectral features are typically not distributed evenly across a spectrum and can have varying degrees of quality.  The strengths of the spectral features can also depend on the variety of objects being studied.

The {\tt feature} mode of \swoc{} was set up to take all of this into account when evaluating the FoM for different combinations of wavelength bands.  This mode needs the following input:
\begin{itemize}
\item a set of line list files for those objects under consideration
\item a list of regions in the spectrum that must be included
\item a list of groups of spectral features that must be included
\item a list of groups of `conditional' spectral features that would be useful/interesting to include.
\end{itemize}

The spectral line list file contains all needed information about the spectral lines and features of interest, including their location, designated name, strengths, and weights depending on their quality.  It is important to generate a line list file for each type of object to be studied.  The program will then iterate through each list, determining the optimal wavelength coverage for each object, and finally, combining all information to ascertain the optimal wavelength coverage for all objects together.

The point of the required wavelength regions and the groups of required spectral features is to set the bottom-line requirements for a given scientific study.  The required wavelength regions consist of any spectral regions that must be included within the combination of wavelength bands in order to carry out the science of a given spectroscopic study.  For example, such regions could include broad absorption features, such as the H$_{\alpha}$ Balmer line (that can be used to measure the effective temperature in cool stars) or diffuse interstellar bands (that trace dust in the interstellar medium), that cover several Ångstr\"{o}ms.   

The groups of required spectral features are those for which at least one instance of each group must be contained within the final combination of wavelength bands.  The program organizes the spectral features provided in the line list file according to each feature's designated name (see Appendix~\ref{sec-infeat}).  Groups of spectral features can thus include one or more type of ion or molecule, or some other categorization devised by the user.  For example, a systematic study of lithium in stars would require lithium lines to be covered by the final combination of wavelength bands.  Thus, the element lithium would be included as one group of required spectral features.  Any combination of wavelength bands that do not satisfy {\it all} requirements are immediately discarded as `failures' in the FoM analysis.

Finally, the list of conditional spectral features consists of those groups that will improve the science outcomes, but the loss of which would not inhibit the science goals of a given spectroscopic study.   For some science cases, there might be zero conditional features. The final FoM is quantified as the number of spectral features in all groups of conditional features present within a given combination of wavelength bands divided by the total number of features across the entire spectral region under study.

\swoc{} computes the FoM using the following steps.  

\begin{itemize}
\item The program first computes a weighted sum of the number of spectral features associated with each groups of conditional features for the spectral range defined by the user:
\begin{equation}
N_{{\rm tot},j} = \sum^n_{i} \frac{1}{w_{i,j}^2}
\label{eq-ntot}
\end{equation}
where $N_{{\rm tot},j}$ is the weighted sum of the spectral features in the $j$th group of conditional spectral features, and $w_{i,j}$ represents a quality weight for the $i$th spectral feature in the $j$th group of conditional spectral features.

\item For each wavelength segment, a vector of bands is created. Each band has a starting wavelength (band head) inside the spectral range defined by the user.  The step-size between subsequent band heads should be chosen according to the required precision needed.  

\item The program then proceeds to define a matrix of all combinations of bands.  Combinations with bands that overlap in wavelength coverage are discarded in order to avoid counting the same spectral feature multiple times.  The program also allows for gaps between the bands. This leaves $k$ distinct combinations of bands.    

\item The program then determines which of the required spectral regions and groups of required spectral features have been satisfied for each combination of bands.  Any combination of bands that does not meet the requirements are automatically discarded.

\item For those combinations of bands that meet the requirements, the program computes a weighted sum of the number of spectral features associated with each group of conditional features, according to:
\begin{equation}
N_{j,k} = \sum^n_{i} \frac{1}{w_{i,j,k}^2}
\label{eq-wsum}
\end{equation}
where $N_{j,k}$ is the weighted sum of the spectral features in the $j$th group of conditional spectral features for the $k$th combination of bands, and $w_{i,j,k}$ represents the quality weight for the $i$th spectral feature in the $j$th group of conditional spectral features.  Individual figures-of-merit (${\rm fom}_{j,k}$) for each group of conditional spectral features are computed by normalizing each $N_{j,k}$ by the maximum value over the entire spectral range considered, $N_{{\rm tot},j}$ (see equation~\ref{eq-ntot}):
\begin{equation}
{\rm fom}_{j,k} = \frac{N_{j,k}}{N_{{\rm tot},j}}
\end{equation}

\item A final figure-of-merit (${\rm FoM}_k$) for each combination of bands is computed by summing over all $j$ groups of conditional spectral features and normalizing by the total number of groups, given as:
\begin{equation}
{\rm FoM}_k = \frac{\sum_1^j {\rm fom}_{j,k}}{N\left({\rm groups of conditional features}\right)}
\end{equation}

\item Those combinations of bands in which ${\rm FoM}_k$ is maximized are selected as the most optimal combinations.

\end{itemize}

The above steps can be completed for a list of objects, each with their own line list file.  Finally, the program determines an overall optimal combination of bands.  This is done by first summing each ${\rm FoM}_k$ across all objects and normalizing by the total number of object line list files to compute a FoM-total.  The combinations of bands for which FoM-total is maximized are considered to provide the optimal spectral coverage for all objects tested.

\subsection{Spectral Mode}\label{sec-spec}

In some studies, we may not want to use (or may not have) a set of spectral features with which to analyze a set of objects.  Instead, we could classify objects according to their overall spectrum.  In this case, we might want to determine in which parts of a spectrum we see the largest variation amongst a group of objects.  This is the idea behind the {\tt spectral} mode in \swoc.  This mode determines the wavelength regions, which show the largest differences among several flux calibrated spectra of objects of interest. In contrast to the {\tt feature} mode, this mode requires very little input from the user.  Thus, the {\tt spectral} mode is considered more objective in nature than the {\tt feature} mode.

The {\tt spectral} mode can be very useful when (1) spectral template fitting is required to categorize objects or to determine some overall parameter of an object, or (2) a specific object must be selected among several different types of objects.  In both cases, it is important to determine the spectral regions that show the greatest variation amongst the spectra for the typical objects that will be observed and identified.  This is exactly what the {\tt spectral} mode achieves.

In this mode, \swoc{} performs the desired difference analysis using the following steps.

\begin{itemize}

\item All spectra are first reduced to the spectral range to be tested, as defined by the user.

\item The square of the difference between the fluxes of each pair of spectra are computed for all wavelength points in the spectra.  For example, if three spectra are listed, then three resultant square difference arrays are computed:
\begin{equation}
\begin{array}{lcc}
\mathbf{D}_{1,2} = \left\{\frac{(f_{1,i} - f_{2,i})^2}{\sigma_{1,i}^2 + \sigma_{2,i}^2}, ...\right\} \\
\mathbf{D}_{1,3} = \left\{\frac{(f_{1,i} - f_{3,i})^2}{\sigma_{1,i}^2 + \sigma_{3,i}^2}, ...\right\} \\
\mathbf{D}_{2,3} = \left\{\frac{(f_{2,i} - f_{3,i})^2}{\sigma_{2,i}^2 + \sigma_{3,i}^2}, ...\right\}
\end{array}
\end{equation}
where $f_i$ is the flux and $\sigma_i$ is the uncertainty at the $i$th wavelength point.  If, rather, the user wants to compare to a reference spectrum, then the square difference arrays will instead be given by:
\begin{equation}
\begin{array}{lcc}
\mathbf{D}_{{\rm ref},1} = \left\{\frac{(f_{1,i} - f_{{\rm ref},i})^2}{\sigma_{{\rm ref},i}^2 + \sigma_{1,i}^2}, ...\right\} \\
\mathbf{D}_{{\rm ref},2} = \left\{\frac{(f_{2,i} - f_{{\rm ref},i})^2}{\sigma_{{\rm ref},i}^2 + \sigma_{2,i}^2}, ...\right\} \\
\mathbf{D}_{{\rm ref},3} = \left\{\frac{(f_{3,i} - f_{{\rm ref},i})^2}{\sigma_{{\rm ref},i}^2 + \sigma_{3,i}^2}, ...\right\}.
\end{array}
\end{equation}
The total square difference array is then computed by vector summing the individual square difference arrays:
\begin{equation}
\begin{array}{lcc}
\mathbf{D}_{\rm TOT} = \mathbf{D}_{1,2} + \mathbf{D}_{1,3} + \mathbf{D}_{2,3}; \\
\mathbf{D}_{\rm TOT} = \mathbf{D}_{{\rm ref},1} + \mathbf{D}_{{\rm ref},2} + \mathbf{D}_{{\rm ref},3}. \\
\end{array}
\end{equation}

\item The program next computes the total difference across the entire spectral range:
\begin{equation}
D_{\rm spec} = \sum_i D_{{\rm TOT},i}
\label{eq-dspec}
\end{equation}
where $D_{{\rm TOT},i}$ is the value of $D_{\rm TOT}$ at the $i$th wavelength point.

\item Similarly to the {\tt feature} mode, a vector of bands is created for each wavelength segment, and a matrix of all combinations of bands is defined (also accounting for any defined gaps between the bands).  This leaves $k$ distinct combinations of bands.  

\item The program then computes a total FoM by summing those elements of $\mathbf{D}_{\rm TOT}$, which correspond to wavelengths contained within the combination of bands, given as:
\begin{equation}
{\rm FoM}_k = \frac{\sum_i D_{{\rm TOT},i}^k}{D_{\rm spec}}
\end{equation}
where $D_{{\rm TOT},i}^k$ is the value of $D_{\rm TOT}$ at the $i$th wavelength point inside the wavelength ranges of the $k$th combination of bands, and $D_{\rm spec}$ is the total difference within the entire spectral range tested (see equation~\ref{eq-dspec}).

\item Those combinations of bands in which ${\rm FoM}_k$ is maximized are selected as the most optimal combinations.  

\end{itemize}

The resultant combination of bands that maximize ${\rm FoM}_k$, as well as $\mathbf{D}_{\rm TOT}$ vs. wavelength, are output by {\tt SWOC}.  Further, the user can request that these be plotted together as an additional output figure.

\section{Feature Mode Example: FGK stars in the Galactic disc and bulge}
\label{sec-mw}

In the following sections, we give several examples to illustrate how \swoc{} works and how to use its two modes.  For more details on the input to \swoc{}, given in the following examples, see Appendix~\ref{sec-inswoc} and Table~\ref{tab-inputs}.

Our first example consists of a survey to study stars in the bulge and disc of the Milky Way.  We assume that our spectrograph will have a resolving power of $R\sim20,000$, which is similar to that of the FLAMES/GIRAFFE spectrograph for the {\it Gaia}-ESO Survey, as well as the planned high-resolution spectrographs for WEAVE and 4MOST.  Since the disc and bulge consist of predominantly metal-rich ($\feh>-1$) stars, we will limit our analysis to spectral features between 4500~\AA{} and 7000~\AA{}.  Below 4500~\AA{}, line blending tends to be an issue in metal-rich stars \citep{hansen2015}.

We first compiled a line list, which contains spectroscopic lines to determine elemental abundances and features that provide diagnostics for the stellar parameters.  A description of the major nucleosynthesis channels included and our chosen line list is given in Appendix~\ref{sec-linelist}.  This list is not necessarily exhaustive, and depending on the science goals for a different survey, one may add or exclude features, and extend the line list to bluer and redder wavelengths. 

\subsection{Determining line strengths}

The strengths of different spectral features depend on the variety of stars being studied.  Thus, it is necessary to include an estimate of the strength of all spectral features.  For example, in the case of a stellar spectroscopic study, synthetic spectra could be computed for the types of stars that will be analyzed.  Using these spectra, theoretical equivalent widths can be measured to determine the strengths of the lines in the spectral feature line list.  Another option is to use the VALD database \citep{kupka1999,kupka2000} to extract line depths for a given stellar type.  Both equivalent widths and line depths are suitable measures of line strength.

In order to measure the strength of the lines, we calculated theoretical equivalent widths (EWs) for the lines in our list.  To measure the line strengths, we used synthetic spectra for typical effective temperatures ($\teff$) and surface gravities ($\logg$) for a main-sequence (MS), turn-off (TO), and red clump (RC) star, with metallicities of $0.0$, $-1.0$, and $-2.0$~dex. The nine different sets of stellar parameters are listed in Table~\ref{tab-syn}.  We set ${\tt feat.min}=20.0$~m\AA.  This means that for a given star/list, all lines weaker than $20$~m\AA{} are considered unmeasurable.  This somewhat emulates the capability of a $R\sim20,000$ spectrograph, which has been used to obtain spectra with signal-to-noise equal to $30-50$~per pixel in the line free continuum.  

\begin{table}
\centering
\caption{Synthetic spectra used for measuring equivalent widths.}
\label{tab-syn}
\begin{threeparttable}
\begin{tabular}{lcccc}
\hline\hline
Star Type & $\teff$ & $\logg$ & $\feh$ & $\Vmic$\tnote{a} \\
& (K) & & & (\kms)\\
\hline
Main-sequence (MS) & 5250 & 4.5 & $(-2,-1,0)$ & 1.0\\ 
Turn-off (TO) & 6250 & 4.0 & $(-2,-1,0)$ &  1.0\\ 
Red Clump (RC) & 4500 & 3.0 & $(-2,-1,0)$ & 2.0 \\ 
\hline\hline
\end{tabular}
\begin{tablenotes}
\item [a] Denotes microturbulence
\end{tablenotes}
\end{threeparttable}
\end{table}

\subsection{Defining the requirements and groups of conditional features}

We have compiled a set of requirements and groups of conditional spectral features, which are listed in Table~\ref{tab-fom}.  We have based our selection of the required and conditional features on a careful investigation of which groups of spectral features provide the best diagnostics for stellar parameters and that cover many of the major nucleosynthesis channels that can be used for Galactic archaeology \citep{freeman2002} and the understanding of the chemical enrichment of the disc and bulge of the Milky Way (see Appendix~\ref{sec-linelist}). Depending on the science of a given survey, it is quite possible that these would differ from those listed in Table~\ref{tab-fom}.

Including more than one indicator for $\teff$ and $\logg$ is necessary to be able to determine stellar parameters for all stellar types (see \S\ref{sec-linesp}).   For our required spectral regions, we thus include the wavelength range that surrounds the ${\rm H}_{\alpha}$ Balmer line, which is an indicator for $\teff$.  We further added ranges which cover either the \nionb{Mg}{b} triplet lines or the \nion{Ca}{I} line at 6162~\AA{} to estimate $\logg$.  See Appendix~\ref{sec-linesp} for more details.

\begin{figure*}
\centering
\includegraphics[height=0.58\textwidth]{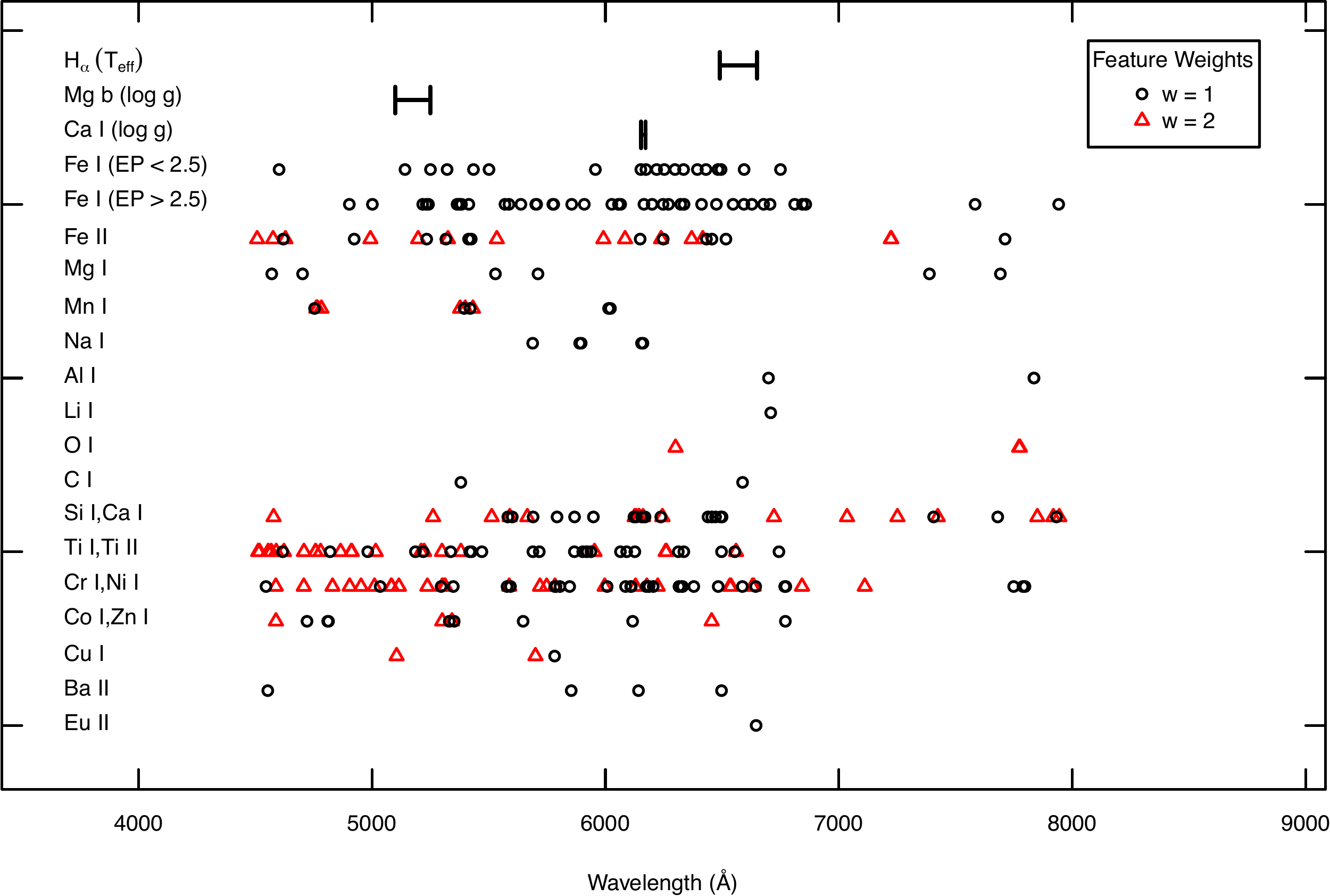}
\caption{Location of spectral features and lines in wavelength.  Elements and features have been grouped according to those listed in Table~\ref{tab-fom}.  The required spectral regions surrounding the strong ${\rm H}_{\alpha}$ Balmer, \nionb{Mg}{b} triplet, and \nion{Ca}{I} lines are shown as black segments. Lines with $w=1$ are shown as black circles, while those with $w=2$ are given by red triangles. }
\label{fig-lines}
\end{figure*}

It is necessary to include a weaker \nion{Mg}{I} or \nion{Ca}{I} line from which to estimate the abundance of each respective element in order to achieve a good estimate of $\logg$ from the broad lines included in the list of required spectral regions (see \S\ref{sec-linesp}).  Thus, we include these in our list of groups of required spectral features.  In addition, a \nion{Mn}{I} line and an \nion{Al}{I} or \nion{Na}{I} line are required since they are extremely important for constraining proton capture and the metallicity-dependent nucleosynthesis of progenitor stars, respectively.  Finally we require at least one neutron capture element indicator, such as a \nion{Ba}{II} or \nion{Eu}{II} line, to constrain heavy element production.  With Mg, Mn, Al or Na, and Ba or Eu we have powerful tracers of the chemical evolution of the Milky Way (see \S\ref{sec-nuc}).  We further impose that only the best quality lines (those with $w=1$), which can potentially give the best results, are used for the majority of the requirements.  However, due to paucity of lines, all lines in the line list can be considered for the \nion{Mn}{I} requirement.  The final requirements are listed in Table~\ref{tab-fom}.

\begin{table}
\centering
\caption{The required and conditional groups of spectral features for the optimization of the wavelength coverage.}
\begin{threeparttable}
\begin{tabular}{llr}
\hline\hline
Group & Ion/Feature & Comment \\
\hline
\multicolumn{3}{c}{Required Spectral Regions} \\
\hline
wreq1\tnote{a} & ${\rm H}_{\alpha}$ & $6470-6650$~\AA \\
wreq2\tnote{b} & \nionb{Mg}{b} & $5100-5250$~\AA \\
           & \nion{Ca}{I} & $6152-6172$~\AA \\
\hline
\multicolumn{3}{c}{Groups of Required Spectral Features} \\
\hline
freq1\tnote{c} & \nion{Mg}{I} & $w=1$ \\
freq2 & \nion{Al}{I}, \nion{Na}{I} & $w=1$ \\
freq3\tnote{c} & \nion{Ca}{I} & $w=1$ \\
freq4 & \nion{Mn}{I} & $w\leq2$ \\
freq5 & \nion{Ba}{II}, \nion{Eu}{II} & $w=1$\\
\hline
\multicolumn{3}{c}{Groups of Conditional Spectral Features} \\
\hline
con1 & \nion{Fe}{I} with $EP\leq2.5$ & $w=1$ \\
con2 & \nion{Fe}{I} with $EP>2.5$ & $w=1$ \\
con3 & \nion{Fe}{II} & $w\leq2$\\
con4 & \nion{Li}{I} & $w\leq2$ \\
con5 & \nion{C}{I} & $w\leq2$\\
con6 & \nion{O}{I} & $w\leq2$\\
con7 & \nion{Na}{I} & $w\leq2$ \\
con8 & \nion{Mg}{I} & $w\leq2$ \\
con9 & \nion{Al}{I} & $w\leq2$ \\
con10 & \nion{Si}{I}, \nion{Ca}{I} & $w\leq2$\\
con11 & \nion{Ti}{I}, \nion{Ti}{II} & $w\leq2$\\
con12 & \nion{Cr}{I}, \nion{Cr}{II}, \nion{Ni}{I} & $w\leq2$\\
con13 & \nion{Co}{I}, \nion{Zn}{I} & $w\leq2$\\
con14 & \nion{Cu}{I} & $w\leq2$\\
con15 & \nion{Ba}{II} & $w\leq2$ \\
con16 & \nion{Eu}{II} & $w\leq2$ \\
\hline\hline
\end{tabular}
\begin{tablenotes}
\item [a] $\teff$ indicator
\item [b] $\logg$ indicators.  Only one is required, and so both would be listed together in the input file.
\item [c] A weak line of the same element as the $\logg$ indicator is required in order to estimate the abundance of that element.  However, we require a weak \nion{Mg}{I} line in either case.
\end{tablenotes}
\end{threeparttable}
\label{tab-fom}
\end{table}

With the requirements in place, we devised a list groups of conditional spectral features, which are also listed in Table~\ref{tab-fom}.  First, we wish to have a statistically significant number of both high and low excitation potential (EP) \nion{Fe}{I} lines, as well as \nion{Fe}{II} lines to successfully determine $\teff$ and $\logg$ through excitation and ionization equilibrium of iron \citep[see Appendix~\ref{sec-linesp} and, e.g.,][]{lind2012}.  Since there are plenty of strong \nion{Fe}{I} lines across a large range of wavelengths, we want to consider only $w=1$ lines.  In practice, there will likely be many more lines that can be used.  However, having a set of highly-weighted lines is desirable.   

The remaining groups of conditional features provide at least one indicator from each nucleosynthesis channel described in \S\ref{sec-nuc}.    Some of the lines for the ions listed can be very weak or suffer from blends or hyperfine splitting.  However, since the program weighs each line according to $1/w^2$ (see eq.~\ref{eq-wsum}), the best lines will dominate the FoM calculations for these ions.    

The location in wavelength of the spectral lines and features, which are used for our requirements and groups of conditional features, is shown in Figure~\ref{fig-lines}.  As can be seen, the lines are not evenly distributed for most elements.  Now that the requirements and groups of conditional features have been selected, wavelength regions can be assessed to determine those that give optimal coverage.

\subsection{Exploring a 2-arm design}\label{sec-2ex}

We now explore the optimal wavelength coverage for several 2-arm spectrograph designs using \swoc.  For simplicity, we here require that both arms must have equal bandwidths.  Although we have chosen this simple setup, our method is easily expanded to accommodate designs with different wavelength limits, a larger set of arms (\S\ref{sec-4ex}), as well as one which has arms of varying wavelength ranges.

For our example 2-arm spectrograph design, we evaluated bands inside 4500 and 8000~\AA{}, which have bandwidths ranging from 550~\AA{} down to 450~\AA{} in steps of 25~\AA.  The band head of each consecutive band was taken in steps of ${\tt bstep}=20$~\AA.  We set ${\tt gap}=0$, as well as ${\tt rv.shift}=10$~\AA{} to account for radial velocities of up to $\sim400$~\kms. The FoM values were then computed with \swoc{} for various combinations of two bands where the bands are always the same width.

\begin{figure}
\centering
\includegraphics[height=0.58\textwidth]{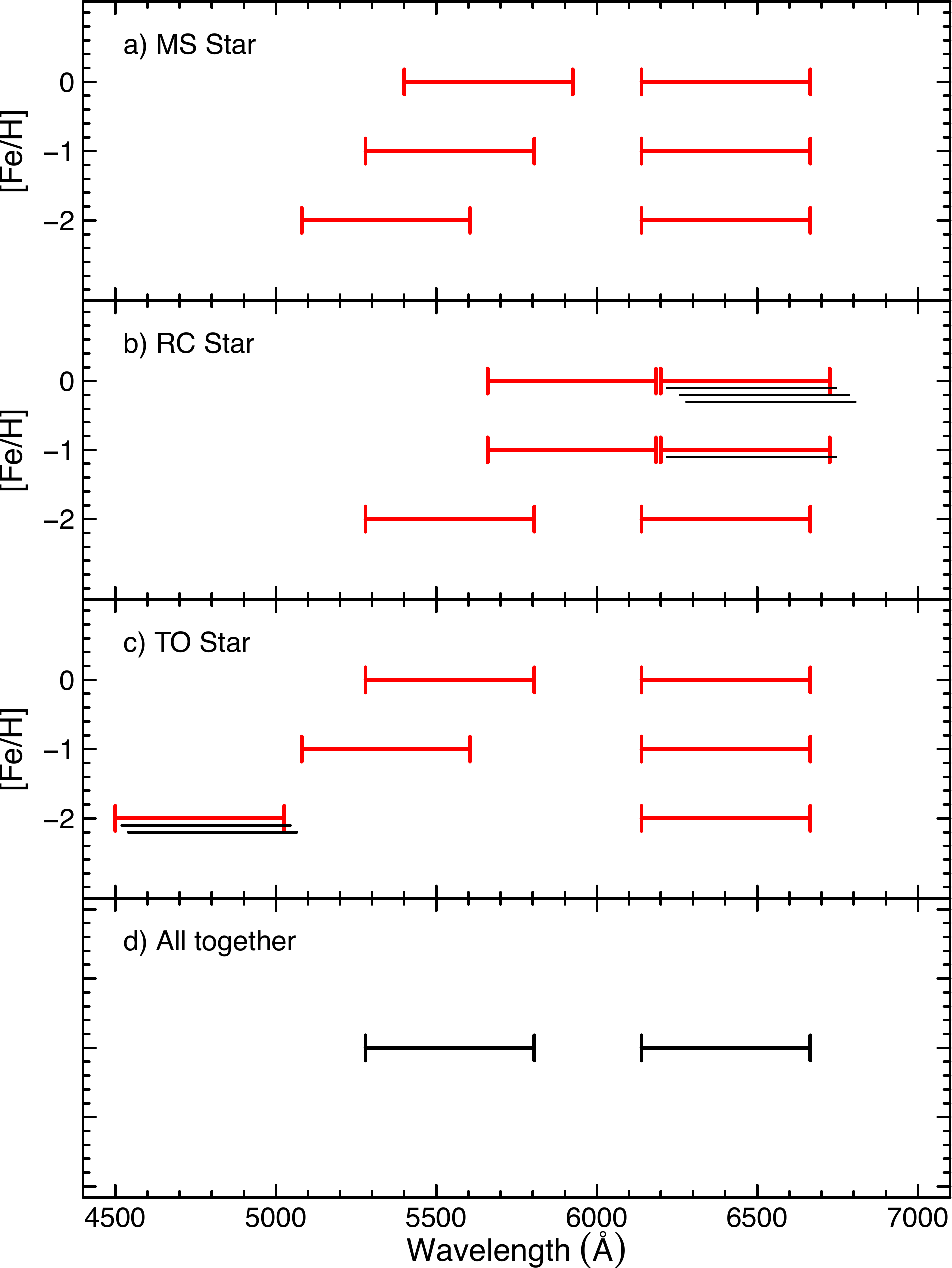}
\caption{Combinations of spectral bands that achieved a maximum FoM in the 525~\AA{} bandwidth case for a) the main-sequence star, b) the red clump star, c) the turn-off star, and d) all stars together.  The stellar parameters for each star are given in Table~\ref{tab-syn}.  The optimal combinations of bands are shown as red segments for each metallicity.  For some metallicities, one of the bands could be combined with several different other bands to achieve a maximum FoM.  These are shown as black segments underneath the red segments.}
\label{fig-ex500}
\end{figure}

The maximum FoM and the band heads of the two bands with that FoM are listed in Table~\ref{tab-scan} for each bandwidth.  In Figure~\ref{fig-ex500}, we show the results for the 525~\AA{} bandwidth case, plotting the optimal band combinations for each individual star (listed in Table~\ref{tab-syn}), as well as the final overall optimal combination of bands.  This illustrates the range of possibilities for each stellar type, but when all information is considered together, a single set of bands provides the optimal coverage.
 
All combinations of bands with bandwidths of 450~\AA{} and smaller failed to meet the requirements listed in Table~\ref{tab-fom}.  In the 475~\AA{} bandwidth case, the optimal combinations contain either the \nionb{Mg}{b} triplet lines or the \nion{Ca}{I} 6162~\AA{} line as the $\logg$ indicator, while the optimal combinations only contained the \nion{Ca}{I} 6162~\AA{} line in the $500-550$~\AA{} bandwidth cases.  In general, the maximum FoM increases with increasing bandwidth.  Note, however, that due to normalization differences, direct comparisons can be misleading. 

It is interesting to point out that in all cases, the dominant \nion{Li}{I} line at 6708~\AA{} is either not covered or is near the edge of the final bands.  The reason for this is that the line is very weak in most stars.  Thus, this line does not contribute to the FoM calculation.  The lithium abundance, however, can provide valuable information about the evolution of stars.  Thus, if the science goals of a survey require such a line, then it would be better handled if \nion{Li}{I} was added as a groups of required spectral features instead of a group of conditional features.

\begin{table}
\centering
\caption{Final FoM results for combinations of two spectral regions of equal bandwidth.}
\label{tab-scan}
\begin{tabular}{cccc}
\hline\hline
Bandwidth & Band1 & Band2 & Max \\
(\AA) & (\AA) & (\AA) & FoM  \\
\hline
550 & 5280 & 6120 & 0.39 \\
 & 5360 & 6120 & 0.39 \\
525 & 5280 & 6140 & 0.37 \\ 
500 & 5680 & 6220 & 0.32 \\ 
475 & 5080 & 6240 & 0.29 \\
& 5080 & 6260 & 0.29 \\ 
& 5080 & 6280 & 0.29 \\
& 5700 & 6200 & 0.29 \\
& 5700 & 6220 & 0.29 \\
& 5700 & 6240 & 0.29 \\
450 & -- & -- & Failed requirements \\ 
\hline\hline
\end{tabular}
\end{table}

\subsection{Exploring a 4-arm design}\label{sec-4ex}

{\tt SWOC} was written to handle any number of bands.  Thus, the analysis need not be limited to two arms.  As an example, we evaluated the combinations of 200~\AA{} wide bands for an example 4-arm spectrograph.  We assume this spectrograph has the same capabilities as that in the 2-arm example.

\begin{figure*}
\centering
\includegraphics[height=0.58\textwidth]{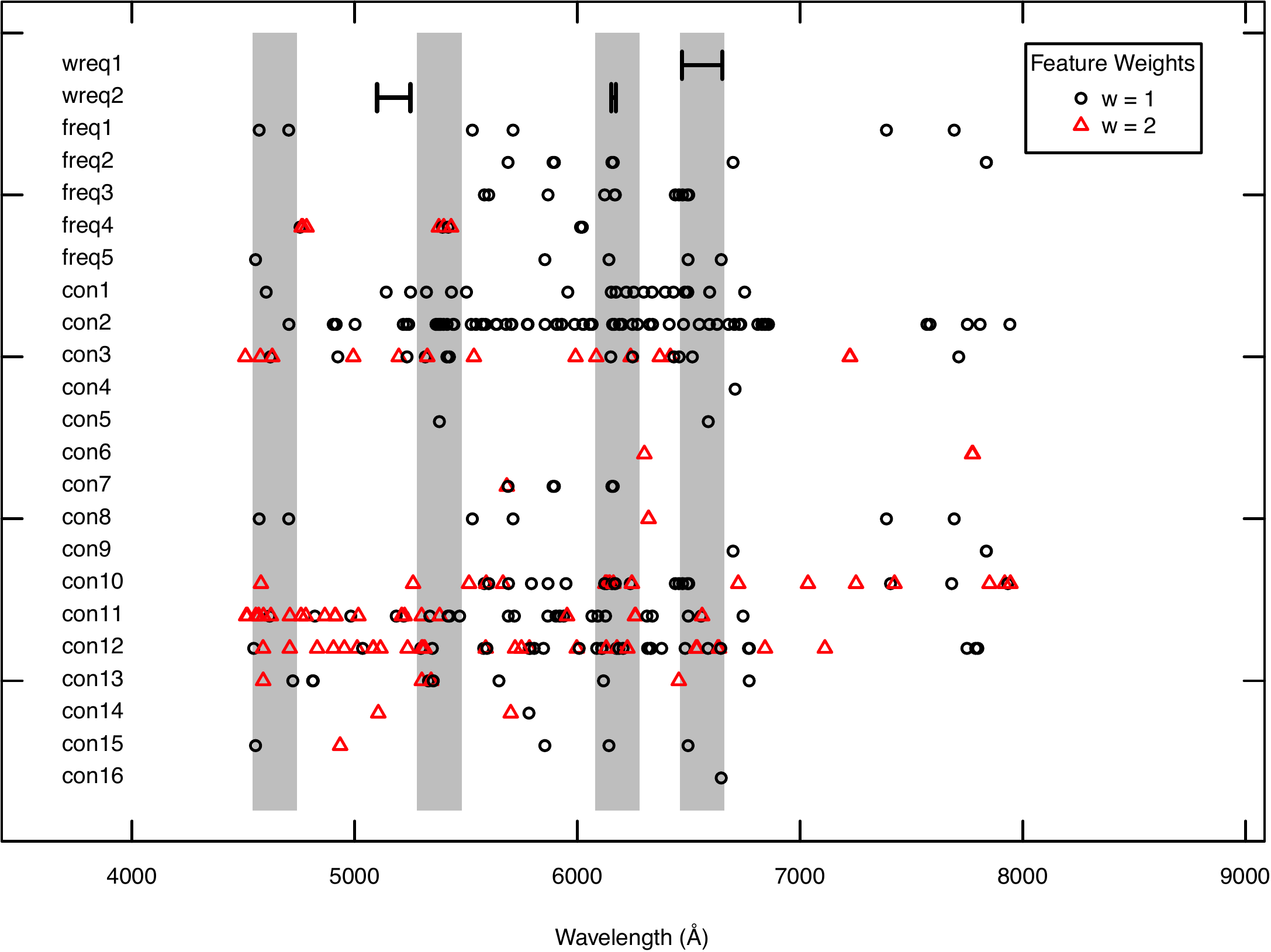}
\caption{Output plot from \swoc{} showing the optimal combination of four 200~\AA{} wide bands for the example 4-arm spectrograph.  The final optimal bands are plotted as grey regions.  The required wavelength regions are shown as black, bracketed line segments.  The spectral features that make up the different groups of required and conditional features (listed in Table~\ref{tab-fom}) are plotted as points, with the symbols and colours corresponding to the weight given to each feature.  The final bands are: $4540-4740$~\AA, $5280-5480$~\AA, $6080-6280$~\AA, and $6460-6660$~\AA.}
\label{fig-4arm}
\end{figure*}

Figure~\ref{fig-4arm} shows the plot that is created by {\tt SWOC} when it has finished running.  Here, the required wavelength regions are shown as black segments, while the points represent the features which contribute to the groups of required and conditional spectral features listed in Table~\ref{tab-fom}, colour-coded according to each feature's weight.  The optimal combination of bands is shown as the grey regions to illustrate what information is available with this configuration.

\subsection{Application to 4MOST}
\label{sec-4most}

The \swoc{} algorithm was originally devised to optimize the wavelength coverage for the green and red arms for the preliminary design of the high-resolution 4MOST spectrograph \citep{dejong2014}.  These regions would be primarily used to perform Galactic archaeology in the disc and bulge of the Milky Way.  The 4MOST spectrograph will also include a blue arm to be used primarily for metal-poor stars in the halo of the Milky Way.  The methods to optimize this blue arm are described in detail in \citet{hansen2015}.

Initially, combinations of two arms restricted to wavelengths between 4500~\AA{} and 8000~\AA{} with varying bandwidths were evaluated with \swoc.  From this analysis, we found that the bandwidths of the two arms must be at least 450~\AA{} wide in order to pass our requirements.  Using this information, the instrument designers found that bandwidths in the green and red arms can be as large as 570~\AA{} and 690~\AA{}, respectively, for potential designs.  These both met our initial recommendations.  However, the designers also required a $\sim300$~\AA{} gap between the two arms due to the cross-over of the dichroic.

\begin{figure*}
\centering
\includegraphics[height=0.58\textwidth]{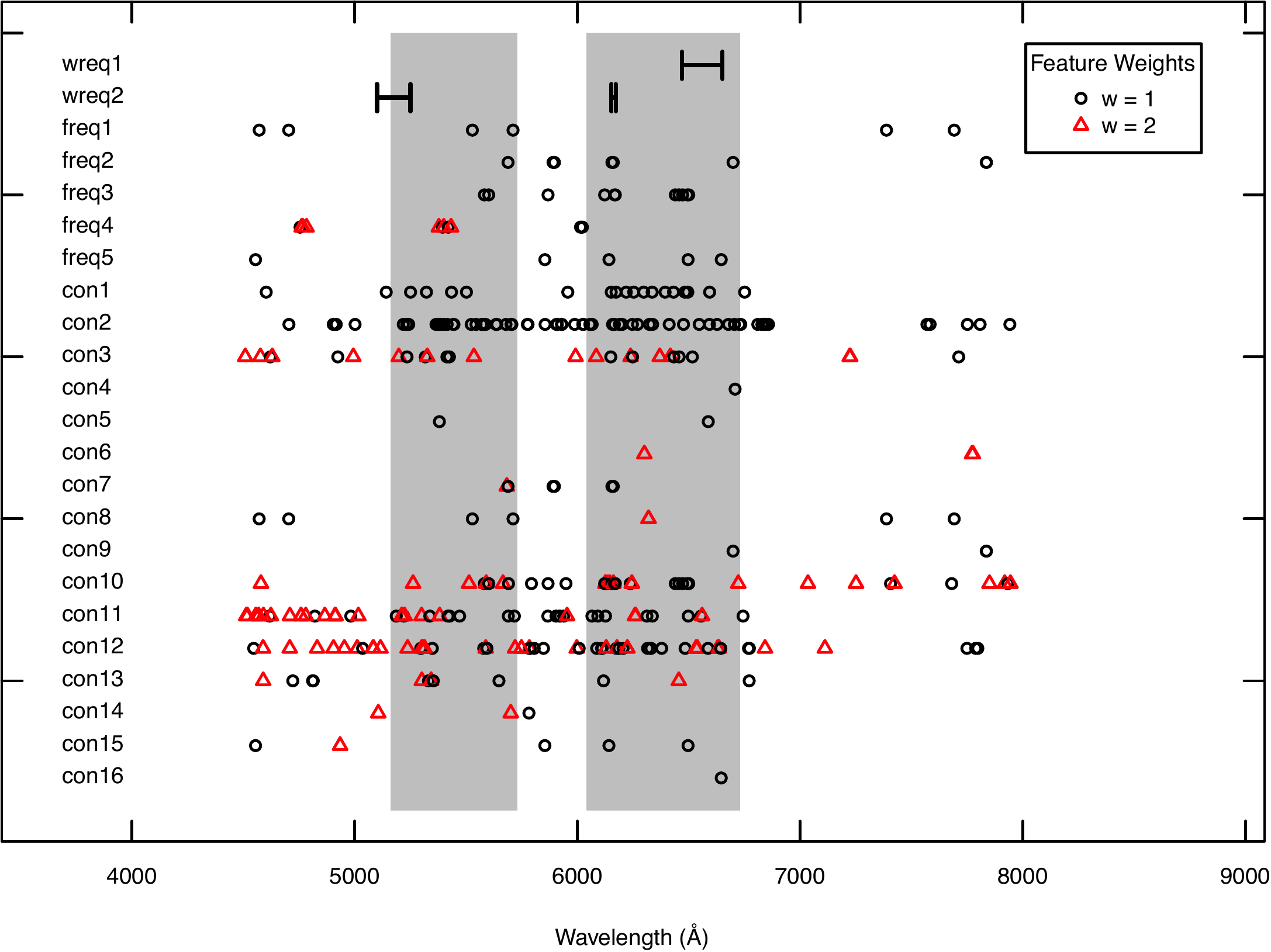}
\caption{Output plot from \swoc{} showing the optimal combination of 570~\AA{} and 690~\AA{} wide bands for the preliminary design for the 4MOST spectrograph.  The final optimal bands are plotted as grey regions, while symbols and colours are the same as defined in Figure~\ref{fig-4arm}.  The final bands chosen are: $5160-5730$~\AA{} and $6040-6730$~\AA.  While this is what came out of \swoc, this is not what is currently implemented in 4MOST because more flexibility was later indicated for the red arm (see \S\ref{sec-4most}). Current design values are $5160-5730$~\AA{} and $6100-6790$~\AA.}
\label{fig-4most}
\end{figure*}

Given this information, we ran \swoc{}, using our defined requirements and conditionals in Table~\ref{tab-fom}, for two arms with bandwidths of 570~\AA{} and 690~\AA{}, respectively.  We imposed a gap, with ${\tt gap}=300$~\AA{}.  Similarly to \S\ref{sec-2ex}, we chose ${\tt bstep}=20$~\AA, ${\tt rv.shift}=10$~\AA, and ${\tt feat.min}=20$~m\AA. This resulted in two optimal combination of wavelength bands: $[5160-5730~\AA{}, 6040-6730~\AA]$ and $[5240-5810~\AA{}, 6120-6810~\AA]$.  Both of these combinations resulted in a ${\rm FoM}=0.41$.  Further assessment showed that the former combination of bands included a large chunk of the \nionb{Mg}{b} triplet (see Figure~\ref{fig-4most}).  Since the \nionb{Mg}{b} triplet is an important feature that can be used to estimate the surface gravity of a star, this set of bands was selected as the final optimal combination.  Note, however, that later analyses at higher precision in {\tt bstep} showed some flexibility in the range for the red arm.  Thus, the current design values for 4MOST are $5160-5730$~\AA{} and $6100-6790$~\AA. 

\section{Spectral Mode Example: choosing a spectral setup}\label{sec-eg}

As described in \S\ref{sec-spec}, the {\tt spectral} mode of {\tt SWOC} can be useful to determine where in a spectrum several objects show the largest spectral differences.  This is especially useful when we want to provide an overall classification of objects.

In this example, we investigated several spectral settings for the FLAMES/GIRAFFE spectrograph \citep{pasquini2002} on the Very Large Telescope at the ESO La Silla Paranal Observatory in Chile.  We utilized the nine stellar spectra listed in Table~\ref{tab-syn}, representing main sequence, turn-off, and red clump stars at three different metallicities (solar, $-1$~dex, and $-2$~dex).  The spectra were all normalized since the {\tt spectral} mode of \swoc{} requires that all spectra are flux calibrated.  We then ran \swoc{} in its static {\tt spectral} mode.  That is, we investigated  specific spectral regions instead of scanning across all possible regions (see Appendix~\ref{sec-static} for more details).  For this example, we investigated only those GIRAFFE settings with wavelengths between 4500~\AA{} and 8000~\AA.  Table~\ref{tab-hr} lists the GIRAFFE settings and their respective wavelength coverage.

Since we want to compare all nine spectra with each other, we set ${\tt ref.spec}=0$ (as explained in Appendix~\ref{sec-inspec}).  \swoc{} will then compute the total difference among the spectra for each GIRAFFE setting and compare this to the total difference among the spectra for the entire spectral range between 4500~\AA{} and 8000~\AA, thus providing a FoM statistic that can be used to assess the distinguishing capability (in terms of overall classification) of each GIRAFFE setting.

The FoM values are also listed for each GIRAFFE setting in Table~\ref{tab-hr}.  The lowest wavelength settings tend to show the largest differences among the spectra.  It appears that HR9A would provide the best spectral range in which we could identify differences and classify the nine spectra.  This setting, as well as the other two settings with ${\rm FoM}\sim0.2$, contains the strong \nionb{Mg}{b} triplet lines.  These lines vary greatly with the surface gravity of a star (see \S\ref{sec-linesp}), and are thus are very useful for classifying the stars.

\begin{table}
\centering
\caption{The {\tt spectral} and {\tt feature} mode FoM results for the FLAMES/GIRAFFE Settings.}
\label{tab-hr}
\begin{tabular}{lccccc}
\hline\hline
Setting & $\lambda$ Start & $\lambda$ End & {\tt spectral} & {\tt feature} \\
 & (nm) & (nm) & FoM & FoM  \\
\hline
HR6 & 453.8 & 475.9 & 0.19 & 0.08 \\
HR7A & 470.0 & 497.4 & 0.17 & 0.05 \\
HR7B & 474.2 & 493.2 & 0.11 & 0.03 \\ 
HR8 & 491.7 & 516.3 & 0.21 & 0.06 \\ 
HR9A & 509.5 & 540.4 & 0.26 & 0.12 \\
HR9B & 514.3 & 535.6 & 0.20 & 0.06 \\ 
HR10 & 533.9 & 561.9 & 0.10 & 0.10 \\
HR11 & 559.7 & 584.0 & 0.04 & 0.07 \\
HR12 & 582.1 & 614.6 & 0.04 & 0.09 \\
HR13 & 612.0 & 640.5 & 0.04 & 0.11 \\
HR14A & 630.8 & 670.1 & 0.04 & 0.09 \\ 
HR14B & 638.3 & 662.6 & 0.03 & 0.06 \\ 
HR15N & 647.0 & 679.0 & 0.02 & 0.06 \\ 
HR15 & 660.7 & 696.5 & 0.01 & 0.02 \\ 
HR16 & 693.7 & 725.0 & 0.02 & 0.00 \\ 
HR17A & 712.9 & 758.7 & 0.03 & 0.01 \\ 
HR17B & 722.5 & 749.0 & 0.02 & 0.00 \\ 
HR18 & 746.8 & 788.9 & 0.01 & 0.00 \\ 
\hline\hline
\end{tabular}
\end{table}

It is important to remember that the {\tt spectral} mode computes the FoM differently from the {\tt feature} mode.  In the {\tt spectral} mode, only differences in the spectra are considered, while the {\tt feature} mode takes into account specific user-defined features in the spectra.  Thus, both modes can result in very different FoM values.  To illustrate this, we computed the FoM in {\tt feature} mode, which is also listed in Table~\ref{tab-hr}.  In this example, we computed the FoM using only the conditionals listed in Table~\ref{tab-fom}.  As shown in Table~\ref{tab-hr}, the FoM values from the two modes can be quite different.

It is interesting to note that the primary GIRAFFE settings used for observations of Milky Way field stars in the {\it Gaia}-ESO Survey were HR10 and HR21.  We do not assess HR21 in our analysis, since it lies outside of the spectral range for our tests.  However, according to the {\tt spectral} mode analysis, HR10 underperforms compared to other settings, such as HR9A and HR9B.  On the other hand, it appears to have a comparable FoM to HR9A (and exceeds that of HR9B) in the {\tt feature} mode analysis.  This is an important distinction since a primary science goal of the {\it Gaia}-ESO Survey is to obtain accurate individual abundances.  Thus, the {\tt feature} mode FoM is more appropriate.

In addition, other considerations must be made beyond just wavelength considerations.  For example, the resolving power of HR10 is $R\sim21,500$, while it is lower for HR9A, at $R\sim18,000$.  Further, the throughput of the HR9A setting is much lower than that of HR10 for the typical objects observed by the {\it Gaia}-ESO Survey.  This means that the exposure times must be longer when using the HR9A setting to achieve the same signal-to-noise values as HR10.  This will lower the survey efficiency.  All of these points together suggest that the HR10 setting is more optimal for the survey science goals.

\section{Conclusions}
\label{sec-con}

In this work, we have presented a new algorithm \swoc, which provides a thorough analysis of the science goals and spectral information to establish the best spectral coverage for a potential spectrograph design or to decide which wavelength settings to use for an existing spectrograph, such as FLAMES \citep{pasquini2002}, HIRES \citep{vogt1994}, or UVES \citep{dekker2000}.  We have described the two modes, {\tt feature} and {\tt spectral}, in which the code operates to evaluate different wavelength configurations and the required input needed for each mode.  The most important features of this analysis are  careful investigations of the useful spectral lines and features ({\tt feature} mode) or the regions of largest variation in a spectrum ({\tt spectral} mode) that are essential for the science requirements for a given spectroscopic study or survey.  With its two modes, the code has been developed so that it can be used for any spectrograph design and survey science goals.

We illustrated the workings of the {\tt feature} mode of \swoc{} by constructing a line list, which bears relevance for Galactic archaeology with FGK stars in the bulge and disc of the Milky Way (see Appendix~\ref{sec-linelist}).  We then ran the program for two example spectrograph designs, showing that it efficiently and effectively determines the best possible combinations of wavelength bands for a set of science requirements.  We also applied our analysis to the preliminary design for the 4MOST spectrograph.  The wavelength range covered by the two redder arms of the high-resolution spectrograph are well defined through our FoM calculations, which include the ability to not only measure elemental abundances but also derive stellar parameters from the spectra themselves.  As discussed in Appendix~\ref{sec-linesp}, this has numerous advantages, including being able to extend the surveys with 4MOST beyond the range of good {\it Gaia} parallaxes and being independent of reddening complications and extinction determinations.  This demonstrates the importance of having well-defined science requirements, in combination with an appraised line list, when determining the wavelength coverage for a given spectrograph.

We also provided an example of the {\tt spectral} mode, which can be used to assess the parts of a spectrum that provide the most distinguishing features among a set of objects.  The {\tt spectral} mode is useful, because it relies on the spectra alone, with very little input by the user.  It can thus provide a relatively objective analysis of how best to classify several objects given a set of spectral settings, as we showed in our example.

\swoc{} has numerous applications across all spectral studies, from stellar spectroscopy to the investigation of high-redshift galaxies.  With the number of spectroscopic surveys coming online over the next decade, and the availability of many instruments at facilities throughout the world, \swoc{} provides a robust and efficient way to assess the spectral coverage that is best suitable to use on such instruments to achieve the desired science goals.

\section*{Acknowledgements}
GRR and SF acknowledge support from the project grant ``The New Milky Way" from the Knut and Alice Wallenberg Foundation.  AJK acknowledges support from the Swedish National Space Board.  CJH was supported by a research grant (VKR023371) from the VILLUM Foundation.

%
\bibliographystyle{mnras}
\bibliography{wave}
%

\appendix

\section{SWOC Input Definitions}\label{sec-inswoc}

All input parameters are read into \swoc{} using an input file, a template for which is provided as part of the program package.  In the following sections, we describe many of the input parameters and their uses.  Many of the input parameters are also defined in Table~\ref{tab-inputs}.  Further details and formatting instructions can be found in the program help files.

\begin{table*}
\centering
\caption{{\tt SWOC} input definitions}
\label{tab-inputs}
\begin{tabular}{ll}
\hline\hline
Input & Definition  \\
\hline
{\tt band.range} & full range of wavelengths over which the FoM analysis will take place\\
{\tt band.static} & list of specific band combinations to be analyzed using the static modulation\\
{\tt band.width} & list of the respective bandwidths for each wavelength segment\\
{\tt bstep} & step size between band heads, needed to create the vector of bands\\
{\tt con} & list of groups of conditional spectral features, consisting of those spectral features used to compute the FoM\\
{\tt feat.min} & minimum line strength, below which lines are considered immeasurable\\
{\tt freq} & list of groups of spectral features required to be present in the final combination of wavelength bands\\
{\tt gap} & minimum gap that must reside between each band of a combination\\
{\tt inFILE} & list of spectral feature line lists ({\tt feature} mode) or list of spectra ({\tt spectral} mode) used for the FoM analysis\\
{\tt nband} & number of wavelength segments\\
{\tt refspec} & reference spectrum with which each spectrum in {\tt inFILE} will be compared in {\tt spectral} mode \\
{\tt rv.shift} & wavelength tolerance to account for possible shifts due to the radial velocity of a given object\\
{\tt spec.sig} & global uncertainty for each spectrum if no variance spectrum is given in the input spectra\\
{\tt wreq} & list of spectral regions required to be present in the final combination of wavelength bands\\
{\tt wunits} & wavelength units, e.g., \AA{} or nm\\

\hline\hline
\end{tabular}
\end{table*}

\subsection{{\tt feature} mode input parameters}\label{sec-infeat}

The full range of wavelengths ({\tt band.range}) over which the FoM analysis will take place should be first determined, as well as the units ({\tt wunits}) of the wavelengths given, which will be used throughout the analysis.  This is purely for aesthetic purposes in the output files and plots.  The number of wavelength segments ({\tt nband}) and their respective bandwidths ({\tt band.width}, e.g. 300, 400, or 500~\AA) can then be defined.  If required, a minimum wavelength gap ({\tt gap}) that must reside between each band of a combination can also be defined.  Finally, the step-size ({\tt bstep}) between band heads is needed to create the vector of bands (see also \S\ref{sec-feat}).

Each spectral feature list should be saved as a text file and listed under {\tt inFILE}.  Each file should include four columns: the spectral feature wavelength center, a designated name for the feature, an estimate of the strength for each feature, and a weight.  The weight parameter must be greater than zero, where the smallest weight is best.  This is used during the FoM analysis either as an exclusion parameter or a weight parameter (see \S\ref{sec-feat}).  Spectral features that are given the same designated name, will be associated with the same required or conditional spectral feature group during the FoM analysis in \swoc.

A minimum line strength ({\tt feat.min}) can be set in the input file to \swoc, below which lines are considered immeasurable.  Further, each spectral feature listed can be given a wavelength tolerance ({\tt rv.shift}) to account for possible shifts due to the radial velocity of a given object.

Next, the required spectral regions should be listed under {\tt wreq}, while the groups of required and conditional spectral features should be listed under {\tt freq} and {\tt con}, respectively. For each group of  spectral features listed in {\tt freq} or {\tt con}, a maximum quality weight should be provided.  Any features associated with that group and that have a quality weight greater than this value will be disregarded in the FoM analysis.

Below is a template for the input file for the example given in \S\ref{sec-2ex}.

\begin{lstlisting}
#INPUT FILE FOR SWOC: FEATURE MODE

#########################
#WAVELENGTH SEGMENT SETUP
#########################

#Wavelength Units
#e.g. nm (nanometer) or Å (Ångstrom)
wunits <- "Å"

#Wavelength Range
#Define full range of lambda in which the FoMs will be computed
band.range <- c(4500.0,8000.0)

#FOR SCANNING MODULATION (swoc.mode=1)
#Number of Bands
#Define the number of wavelength segments to be considered
nband <- 2

#Bandwidths
#Define the size of each wavelength segment
band.width <- c(500.0,500.0)

#Band gap
#Define the size of the wavelength gap between each wavelength segment
gap <- 0

#Step Size
#Define the step size to determine the bands
bstep <- 20

#FOR STATIC MODULATION (swoc.mode=2)
#List bands / band combinations for which a FoM will be computed.
#Only used if swoc.mode=2 in main swoc program
#band.static[[1]] <- rbind(c(5160,5730),c(6040,6730))
#band.static[[2]] <- rbind(c(5200,6300))

########################################
#LOAD FEATURE FILES AND INITIALIZE SETUP
########################################

#Directory containing line list files
#e.g. FILE.dir <- "swoc/spec/"
FILE.dir <- paste(.libPaths(),"/swoc/extdata/",sep="") #Path to run example files that come with the SWOC package

#List line list files to be considered
#Files should contain the following: Lambda, feature name, strength (e.g., EW, line depth, etc.), flag
#Files can be zipped or unzipped.
inFILE <- c(
	"MSp0.0.xy_47K_full.4m.ldat.gz",
	"MSm1.0.xy_47K_full.4m.ldat.gz",
	"MSm2.0.xy_47K_full.4m.ldat.gz",
	"RCp0.0.xy_47K_full.4m.ldat.gz",
	"RCm1.0.xy_47K_full.4m.ldat.gz",
	"RCm2.0.xy_47K_full.4m.ldat.gz",
	"TOp0.0.xy_47K_full.4m.ldat.gz",
	"TOm1.0.xy_47K_full.4m.ldat.gz",
	"TOm2.0.xy_47K_full.4m.ldat.gz"
	)

#Line wavelength shifts
#Wavelength shift added and subtracted from line centers to account for RV shifts
rv.shift <- 10.0

#Minimum feature strength
#Define the minimum freature strength, which sets what features are strong enough to "measure".
feat.min <- 20.0

#########################
#FIGURE-OF-MERIT SETUP
#########################

#Required Spectral Regions
#Useful for broad features, or groups of features that are required no matter what
#Denote in Å
#wreq <- NULL #Set if no wavelength requirements
wreq[[1]] <- rbind(c(6470.0,6650.0))
wreq[[2]] <- rbind(c(5100.0,5250.0),c(6152.0,6172.0))

#Groups of Required Spectral Features
#List spectral features or groups of spectral features required regardless of strength
#freq <- NULL #Set if no element requirements
freq[[1]] <- rbind(c(12.0,1))
freq[[2]] <- rbind(c(11.0,1),c(13.0,1))
freq[[3]] <- rbind(c(20.0,1))
freq[[4]] <- rbind(c(25.0,2))
freq[[5]] <- rbind(c(56.1,1),c(63.1,1))
freq[[7]] <- rbind(c(6.0,1))

#Groups of Conditional Features
#A Figure-of-Merit is computed based on spectral feature strengths (e.g., equivalent widths, depth)
#List spectral features or groups of spectral features desired according to name and the maximum weight.
#each feature given as c(name,weight), e.g. for Mg lines with weight <= 2: c(12.0,2)
con[[1]] <- rbind(c(26.0,1)) #FeI, EP < 2.5
con[[2]] <- rbind(c(26.25,1)) #FeI, EP >= 2.5
con[[3]] <- rbind(c(26.1,2)) #FeII
con[[4]]  <- rbind(c(3.0,2)) #LiI
con[[5]]  <- rbind(c(6.0,2)) #CI
con[[6]]  <- rbind(c(8.0,2)) #OI
con[[7]]  <- rbind(c(11.0,2)) #NaI
con[[8]]  <- rbind(c(12.0,2)) #MgI
con[[9]]  <- rbind(c(13.0,2)) #AlI
con[[10]]  <- rbind(c(14.0,2),c(20.0,2)) #SiI, CaI
con[[11]]  <- rbind(c(22.0,2),c(22.1,2)) #TiI, TiII
con[[12]]  <- rbind(c(24.0,2),c(24.1,2),c(28.0,2)) #CrI, CrII, NiI
con[[13]] <- rbind(c(27.0,2),c(30.0,2)) #CoI, ZnI
con[[14]] <- rbind(c(29.0,2)) #CuI
con[[15]] <- rbind(c(56.1,2)) #BaII
con[[16]] <- rbind(c(63.1,2)) #EuII

#########################
#OUTPUT FILES
#########################

#Where do you want results to be output?
outdata <- "out_test.txt"

#Would you like the results plotted? TRUE/FALSE
#If TRUE, resultant plots will be saved as a postscript file
plot.results <- TRUE
#Where would you like the plots saved?
outplot <- "out_test.ps"
\end{lstlisting}

\subsection{{\tt spectral} mode input parameters}\label{sec-inspec}

The input parameters for the {\tt spectral} mode in \swoc{} very much resemble those defined in {\tt feature} mode.  All wavelength segment parameters should be defined as described in \S\ref{sec-infeat}.  In this mode, the spectra that will be analyzed should be listed in {\tt inFILE}.  The spectra should be in text format with two or three columns, the first being the wavelength and the second, flux.  The third column is optional. If included, it should be the variance in the flux for each point in the spectrum.  If no variance spectrum is included, then a global uncertainty value should be supplied under {\tt spec.sig} for each spectrum listed in {\tt inFILE}.  If only a single value is given, this uncertainty is assumed for all spectra.  If the user wishes to ignore uncertainties (highly unadvised if actual observations are being used), then set ${\tt spec.sig}=0$.

In addition, the user can supply a reference spectrum under {\tt ref.spec}.  In this case, the spectra listed in {\tt inFILE} will be compared with this spectrum only and not each other.  This is useful if the user is trying to determine the wavelength regions that could be best used to select an object with a typical spectrum like that in 
{\tt ref.spec} amongst objects that will have typical spectra like those listed in {\tt inFILE}.  If the user wishes to not use a reference spectrum, and instead compute the differences among all spectra in {\tt inFILE}, then just set ${\tt ref.spec}=0$.

\swoc{} computes direct differences between each spectrum.  It is therefore important that the spectra are already flux and continuum normalized (if not synthetic spectra), and also have the same sampling (i.e. the wavelength points are consistent among all spectra).  No additional input is needed from the user.  

Below, we provide an example input file for the test described in \S\ref{sec-eg}.

\begin{lstlisting}
#INPUT FILE FOR SWOC: SPECTRAL MODE

#########################
#WAVELENGTH SEGMENT SETUP
#########################

#Wavelength Units
#e.g. nm (nanometer) or Å (Ångstrom)
wunits <- "nm"

#Wavelength Range
#Define full range of lambda in which the FoMs will be computed
band.range <- c(450.0,800.0)

#FOR SCANNING MODULATION (swoc.mode=3)
#Number of Bands
#Define the number of wavelength segments to be considered
nband <- 1

#Bandwidths
#Define the size of each wavelength segment
#band.width <- c(45.0,45.0)

#Band gap
#Define the size of the wavelength gap between each wavelength segment
gap <- 10.0

#Step Size
#Define the step size to determine the bands
bstep <- 2

#FOR STATIC MODULATION (swoc.mode=4)
#List bands / band combinations for which a FoM will be computed.
#Only used if swoc.mode=2 in main swoc program
band.static[[1]] <- rbind(c(453.8,475.9)) #HR6
band.static[[2]] <- rbind(c(470.0,497.4)) #HR7A
band.static[[3]] <- rbind(c(474.2,493.2)) #HR7B
band.static[[4]] <- rbind(c(491.7,516.3)) #HR8
band.static[[5]] <- rbind(c(509.5,540.4)) #HR9A
band.static[[6]] <- rbind(c(514.3,535.6)) #HR9B
band.static[[7]] <- rbind(c(533.9,561.9)) #HR10
band.static[[8]] <- rbind(c(559.7,584.0)) #HR11
band.static[[9]] <- rbind(c(582.1,614.6)) #HR12
band.static[[10]] <- rbind(c(612.0,640.5)) #HR13
band.static[[11]] <- rbind(c(630.8,670.1)) #HR14A
band.static[[12]] <- rbind(c(638.3,662.6)) #HR14B
band.static[[13]] <- rbind(c(647.0,679.0)) #HR15N
band.static[[14]] <- rbind(c(660.7,696.5)) #HR15
band.static[[15]] <- rbind(c(693.7,725.0)) #HR16
band.static[[16]] <- rbind(c(712.9,758.7)) #HR17A
band.static[[17]] <- rbind(c(722.5,749.0)) #HR17B
band.static[[18]] <- rbind(c(746.8,788.9)) #HR18

#########################
#LOAD SPECTRAL FILES
#########################

#Directory containing line list files
#e.g. FILE.dir <- "swoc/spec/"
FILE.dir <- paste(.libPaths(),"/swoc/extdata/",sep="") #Path to run example files that come with the SWOC package

#List spectral files to be considered
#Files should contain the following: Lambda, flux, variance (optional)
#All input spectral data should already be flux normalized to the same scale.
#Files can be zipped or unzipped.
inFILE <- c(
	"MSp0.0.xy_20K_full.dat.gz",
	"MSm1.0.xy_20K_full.dat.gz",
	"MSm2.0.xy_20K_full.dat.gz",
	"RCp0.0.xy_20K_full.dat.gz",
	"RCm1.0.xy_20K_full.dat.gz",
	"RCm2.0.xy_20K_full.dat.gz",
	"TOp0.0.xy_20K_full.dat.gz",
	"TOm1.0.xy_20K_full.dat.gz",
	"TOm2.0.xy_20K_full.dat.gz"
	)

#If variance not provided, global uncertainty should be given.
#Can be single value, or vector with length equal to inFILE.
#Set spec.sig <- 0 to ignore uncertainty.
spec.sig <- 0

#List reference spectrum to which spectra in inFILE will be compared.
#File should contain the following: Lambda, flux, variance (optional)
#Set ref.spec <- 0 if no reference spectrum.
#ref.spec <- "RCm1.0.xy_20K_full.dat.gz"
ref.spec <- 0
#If variance not provided, global uncertainty should be given.
#Set ref.sig <- 0 to ignore uncertainty.
ref.sig <- 0

#########################
#OUTPUT FILES
#########################

#Where do you want results to be output?
outdata <- "outspec_test.txt"

#Would you like the results plotted?
#If TRUE, resultant plots will be saved as a postscript file
plot.results <- TRUE
#Where would you like the plots saved?
outplot <- "outspec_test.ps"
\end{lstlisting}

\subsection{The static modulation}\label{sec-static}

Instead of scanning through a range of possible wavelength bands, \swoc{} can also be used to assess the figures-of-merit for specific wavelength bands and combinations.  This is know as the ``static" modulation and can be performed in both {\tt feature} and {\tt spectral} modes.  Both modes run similarly to their scanning modulations, but in this case, the user lists the specific bands and combination of bands under {\tt band.static} in the inpute file.  The final output will consist of the defined wavelength bands or combination of bands and their respective computed figures-of-merit.  These can then be compared to assess the performance of different setups.

\section{A line list for FGK stars}
\label{sec-linelist}

Here, we describe the elements and spectral features (between $4500$~\AA{} and $7000$~\AA) that are useful when conducting a survey of FGK stars in the bulge and disc of the Milky Way.  The final line list (described in \S\ref{sec-finallist}) contains many spectroscopic lines and features that provide diagnostics for the stellar parameters, and that cover many of the major nucleosynthetic channels.  This list is not meant to be exhaustive, and depending on the science goals for a given survey and the resolution of the spectrograph, one may add or exclude elements, and extend the line list to bluer and redder wavelengths.

\subsection{Determination of stellar parameters}
\label{sec-linesp}

Stars in the disc and the bulge will be moderately to heavily affected by reddening. This means that we cannot exclusively rely on photometric colours for the derivation of effective temperatures ($\teff$). Also surface gravities ($\logg$) need independent confirmation by spectroscopy since the astrometric inference of $\logg$ may suffer from considerable uncertainties stemming from uncertainties in mass, when studying populations with a wide age range. The inclusion of a measure of $\logg$ will also allow us to extend a survey to distances where the parallaxes are poor (even if such an approach might need calibration, see the discussion in \citealt{bensby2014} and notes in \citealt{nissen2014}) or nonexistent. From these points of view it is highly desirable to include spectroscopic measures of both $\teff$ and $\logg$ directly in the selected wavelength regions.

Often spectroscopic analyses make use of ionization equilibrium for the determination of $\logg$ and excitation balance for $\teff$ \citep[e.g.][]{fulbright2000,adibekyan2012,bensby2014}. Ionization equilibrium from neutral and singly ionized Fe (\nion{Fe}{I} and \nion{Fe}{II}) is often used to derive the surface gravity but also \nion{Ca}{I} and \nion{Ca}{II} lines (e.g. in the {\it Gaia}-ESO Survey) and \nion{Ti}{I} and \nion{Ti}{II} lines (e.g. GALAH) can be used.

The main stumbling block with this technique for a survey with limited wavelength coverage is usually the paucity of unblended and strong lines from the singly ionized ions (e.g. \nion{Fe}{II}). However, also high-resolution studies with wide coverage have reported problems with ionization balance as a measure of surface gravity (see, e.g. \citealt{bensby2014} and references therein and the plots in \citealt{anders2014} for APOGEE). In particular, the main sequence just below the turn-off and further down suffer from increasingly underestimated $\logg$ with lower $\teff$, resulting in an unphysical morphology for the main sequence. The red giant branch is similarly affected, but since the direction of the stellar sequence here aligns with the direction in which the parameters are skewed, the effect is less obvious, but still present.

Studies have shown that 1-D model atmospheres in combination with the assumption of local thermodynamic equilibrium (LTE) are likely the cause of these problems \citep[e.g.][]{mashonkina2007,bergemann2012nlte}.  Indeed, \citet{ruchti2013} illustrated how, at low metallicities, inclusion of non-LTE (NLTE) corrections for \nion{Fe}{I} \citep[from][]{lind2012} can strongly mediate the problem with underestimated $\logg$ from ionization balance on the red giant branch. It is thus likely that inclusion of NLTE effects, possibly in combination with 3D atmospheric models \citep[e.g.][]{ludwig2009,magic2013}, in the analysis will help, however, it remains to be shown that this can fully solve the problem with traditional spectroscopic parameters at all metallicities.

A robust complement to excitation and ionization balance is the inclusion of broad features whose shape, in particular their extending wings, serve as sensitive diagnostics of stellar parameters. Suitable examples for FGK stars are the \nion{Ca}{II} near infrared triplet at 8500~\AA, the optical \nionb{Mg}{b} lines at 5170~\AA{} and the \nion{Ca}{I} line at 6162~\AA. The \nionb{Mg}{b} lines and the \nion{Ca}{I} line at 6162~\AA{} are all strong except at very low metallicities ($\feh<-2.5$) and their wings are affected by pressure broadening and can thus be used to determine $\logg$ \citep[e.g.][]{edvardsson1988,fuhrmann1998,feltzing1998,caffau2013}.  Similarly, the wings of the ${\rm H}_{\alpha}$ and ${\rm H}_{\beta}$ lines are mostly sensitive to effective temperature ($\teff$) in cool stars, via the excitation properties of neutral hydrogen. We have therefore included broad hydrogen and metal lines in our list (see Table~\ref{tab-sp}).

A few remarks are in order. Firstly, the two Balmer lines have different advantages and disadvantages connected to their different sensitivities to stellar parameters and convection treatment. We prefer ${\rm H}_{\alpha}$ mainly because it is situated in a region that is much less crowded with blending metal lines. Hence it is easier to recognize and model the shape of the line (i.e. it will work well also at moderate resolution). Secondly, we note that both Balmer lines lose sensitivity to effective temperature in cooler stars, and are therefore not useful below $\sim4500$~K. Thus, they cannot serve exclusively as spectroscopic indicators of $\teff$. Thirdly, it is important to realize that the usefulness of broad lines of Mg and Ca for surface gravity determination is dependent on the knowledge of their respective abundances. Other, weaker lines of the same element must simultaneously provide information about these. This requirement will be important when we define wavelength regions (see \S\ref{sec-mw}).

We have compiled a list (see Table~\ref{tab-sp}) that includes the most important features for determination of stellar parameters from the spectra. These should be considered when discussing the wavelength coverage.  Note, that this list does not take ionization balance into account.  This is instead included in the full line list (see also discussion in \S\ref{sec-mw} and Table~\ref{tab-fom}).

\begin{table}
\centering
\caption{List of lines and spectral features that are of high relevance for the determination of $\teff$ and $\logg$.}
\begin{threeparttable}
\begin{tabular}{cccc}
\hline\hline
$\lambda$ (\AA) & Line/Feature & $w$ & star flag \\
\hline
\multicolumn{4}{c}{Effective temperature ($\teff$)} \\
\hline
4861.35 & ${\rm H}_{\beta}$ & 2 & D, RC \\
6562.83 & ${\rm H}_{\alpha}$ & 1 & D, RC \\
\hline
\multicolumn{4}{c}{Surface gravity ($\logg$)} \\
\hline
5167.32 & \nionb{Mg}{b} & 1 & D \\
5172.68 & \nionb{Mg}{b} & 1 & D \\
5183.60 & \nionb{Mg}{b} & 1 & D \\
6162.17 & \nion{Ca}{I} & 1 & D, RC \\
\hline
\multicolumn{4}{c}{Interstellar reddening\tnote{a}} \\
\hline
5889.95 & \nion{Na}{D} & 2 & D, RC \\
5895.92 & \nion{Na}{D} & 2 & D, RC \\
\hline\hline
\end{tabular}
\begin{tablenotes}
\item [a] See \S\ref{sec-oddz}.
\end{tablenotes}
\end{threeparttable}
\label{tab-sp}
\end{table}

\subsection{Nucleosynthesis channels}
\label{sec-nuc}

In the following sections, we highlight some of the most important elements that probe different nucleosynthesis paths.  This list is not meant to be complete, but offers a concise description of those elements which have been shown to be useful for chemical evolution models, as well as the feasibility of observing such elements.

For some of the most important elements that probe the different nucleosynthetic paths there is only one or a few lines available in the optical and near infrared spectrum. These lines are hence of high importance when choosing the wavelength range(s). Finally, for those elements with many lines available we wish to have as many lines as feasible and also ensure that we include lines that are strong and un-blended enough to be analyzed in a wide range of stars, e.g. metal-rich dwarf and red clump stars as well as stars at low metallicities ($\feh\sim-2$).

\subsubsection{Lithium}

Lithium has many unique properties and a mixed origin, including Big Bang, stellar, and interstellar medium (ISM) nucleosynthesis.  The low stellar abundances of Li are disproportionate to the immense interest in this light element.  It is an important diagnostic tool for cosmology \citep[e.g.][and references therein]{fields2011}.  In particular, the connection to Big Bang nucleosynthesis has prompted many studies of Li in metal-poor stars \citep[e.g.][]{spite1982,bonifacio2007,sbordone2010}.  Further, Li can also serve as an indicator of age \citep[e.g.][]{donascimento2009,castro2011} and the internal physics of stars \citep[e.g.][]{wallerstein1969,gustafsson1989,korn2006,ruchti2011li} of all metallicities, because the fragile nuclei are destroyed in the hot stellar interiors.  Lithium is also important for constraining cluster formation \cite[e.g.][]{lind2009b}.  In medium-resolution surveys, the element can only be studied from a single line at 6708~\AA.

\subsubsection{Carbon and Nitrogen}

Carbon and nitrogen are primarily produced in AGB stars.  Thus, the abundances of these elements are important for constraining stellar evolutionary models of low-mass stars \citep[e.g.][]{chiappini2003b}.  Further, the abundances of these elements can be used to probe the nucleosynthesis and internal mixing processes that take place between the interior and surface of low-mass giant stars \citep[e.g.][and references therein]{karakas2014}.

Many stars, especially at low metallicity ($\feh\leq-2$), have been observed to have large enhancements in [C/Fe] \citep[e.g.][]{cohen2005,beers2005araa,lucatello2006}.  Studies of these carbon enhanced metal-poor (CEMP) stars have primarily focused on determining their frequency and understanding the enrichment processes that took place to cause these stars to have such large C abundances.  For example, one possibility is mass transfer from a binary companion \citep[e.g.][]{suda2004}. 

The [C/O] ratio in the Galactic discs has also been under debate and is highly relevant, for example, for understanding planet formation and the possible existence of terrestrial planets \citep[e.g.][]{nissen2013}.

There are only two useful atomic \nion{C}{I} lines between 4500~\AA{} and 8000~\AA, at 5380.3~\AA{} and 6587.6~\AA.  These lines, however, are quite weak.  On the other hand, there are no useful atomic \nion{N}{I} lines between these wavelengths.  Several molecular features exist that can be used to constrain the abundances of C and N.  For example, there is the C$_2$ Swan band at 5135.6~\AA{}, as well as a weak CN feature at 6332.2~\AA{}.  We have not included these features in the line list used to illustrate our method.  However, they could readily be included if the science so demands, e.g. when studying CEMP stars.   There are other, stronger, features at wavelengths bluer and redder than our set range, such as the CH G-band, which can further provide diagnostics \citep[e.g.][]{hansen2015}.  Again, it is easy to extend our methodology to bluer and redder wavelengths as needed.

\subsubsection{The $\alpha$-elements: O, Mg, Si, Ca}

The $\alpha$-elements all share a similar behavior with metallicity, characterized by an enhanced plateau up to $\feh\sim-0.5$~dex where a gradual decrease sets in \citep[e.g.][]{fuhrmann1998,fulbright2002,adibekyan2012,bensby2014}. This is traditionally explained by the dominance of core-collapse supernovae, such as Type~II supernovae (SNe~II) and hypernovae (HNe), in the early universe and the delayed onset of Type~Ia supernovae (SNe~Ia).   Thus, the $\alpha$-elements can serve as tracers of the star formation rate and star formation efficiency within a stellar population \citep[e.g.][]{matteucci1990}.  The $\alpha$-elements are also among the best elements with which we can recognize accreted stars in the Milky Way \citep[e.g.][]{tolstoy2009, nissen2010}.

Observationally, O, Mg, and Si have been found to show a clear separation between the `canonical' thin and thick discs, while the separation is less clear for Ca \citep[e.g.][]{bensby2014}.  However, Si is overproduced with respect to O and Mg in pair-instability supernovae (PISN), and thus the ratio of these $\alpha$-elements with respect to each other can serve to trace PISN progenitors.  Along with the other light elements ($Z\leq13$), O and Mg also display anomalies in globular clusters and can be used to study the globular cluster/halo field connection \citep[e.g.][]{gratton2012}.

There are several good Si and Ca lines above $\sim5500$~\AA.  However, Si lines become very weak for metal-poor stars.  On the other hand, there are only a handful of good weak Mg lines.  O can only be constrained using the oxygen triplet near 7770~\AA{} and the forbidden line at 6300~\AA.  However, the forbidden line is often very weak and can be contaminated by telluric lines.   

\subsubsection{Titanium}

Titanium displays an $\alpha$-element like behavior with metallicity and is one of the clearest separators of the thin and thick disc \citep[e.g.][]{bensby2014}.  However, models of Galactic chemical evolution currently fall short of explaining the [Ti/Fe] enhancement at low metallicities.  This may be resolved with multidimensional nucleosynthesis calculations of SNe yields \citep{maeda2003,nomoto2013}.  

Spectra of cool stars have several excellent lines from the neutral atom.  However, they span a short range in excitation potential. Lines of the singly ionized ion are considerably fewer and often weak.

\subsubsection{Light odd-$Z$ elements: Na, Al, K}
\label{sec-oddz}

Contrary to the $\alpha$-elements, Na and Al ratios with respect to Fe in field stars increase up to $\feh\sim-1.0$, because their nucleo-synthesis yields strongly depend on the metallicity of the progenitor stars \citep{kobayashi2006}. [Na/Fe] and [Al/Fe] thereafter turn over and decrease down to solar metallicity as SNe~Ia contribute Fe but no Na and Al.  As with O and Mg, Na and Al both show anomalies in globular clusters \citep[e.g.][]{lind2011b,carretta2013, gratton2012}.  On the other hand, the chemical evolution of potassium is poorly studied as it is only accessible through one line \citep[e.g.][]{carretta2013K}.

There are several good Al and Na  lines within our wavelength range, however, they are all of weak to moderate strength at solar metallicity.  Thus, they are quite weak in metal-poor stars.  In addition, the strong \nion{Na}{D} lines, listed in Table~\ref{tab-sp}, can be used for measurements of interstellar reddening \citep[e.g.][]{munari1997}.  Further, in regions with low extinction and for stars with sufficiently large radial velocities, they can be used as abundance indicators for metal-poor stars.

\subsubsection{Iron-peak elements: Sc, V, Cr, Mn, Co, Ni, Cu, Zn}

Observationally, Sc and V approximately behave like $\alpha$-elements in the disc \citep[e.g.][]{adibekyan2012} but are underproduced by chemical evolution models \citep[e.g.][]{kobayashi2006}.  Mn is extremely important for characterizing SNe~Ia enrichment \citep[e.g.][]{seitenzahl2013}.  Contrary to many elements, SNe~Ia raise the [Mn/Fe] ratio \citep[e.g.][]{francois2004}. In addition, the yields from SNe~Ia may be metallicity dependent \citep[e.g.][]{mcwilliam2003,kobayashi2009, travaglio2011}.  Cr and Ni constrain the mixing efficiency and fallback times of SN ejecta at early times, while they later follow Fe \citep[e.g.][]{mcwilliam1995,bensby2014}. Ni shows an interesting upturn at super-solar metallicities \citep[e.g.][]{pompeia2003,bensby2014} and an interesting split between the low and high-$\alpha$ component of the local halo \citep{nissen2010}, which most likely comes from the fact that low-$\alpha$ stars are likely born in satellite galaxies and accreted into the halo \citep[e.g.][]{tolstoy2009}.  

Zn and Co, on the other hand, constrain the explosion energy of core-collapse SNe \citep[e.g.][]{kobayashi2006,ohkubo2006}, with Zn providing the strongest constraints.  Finally, Cu reacts to both metallicity (like Na) and energy (like Zn). The [Cu/Fe] ratio shows an increasing trend with metallicity that does not turn-over at $\feh\sim-1.0$, because SNe~Ia yield more Cu than Fe \citep[e.g.][]{matteucci1993,reddy2006}.

Both Sc and V have several lines throughout the optical/near-infrared spectrum, but only \nion{Sc}{II} lines are visible in metal-poor stars.  Many lines of neutral Cr, Ni, and Mn are available. However, not all lines perform equally, and there are unidentified blends in some Mn lines.  Zn has only a handful of lines in a limited wavelength region, which are also affected by continuum issues. There are plenty of Co lines in the optical range, however, they are mostly quite weak.  Unfortunately, there are few useful Cu lines and these tend to be contaminated by molecular lines, which are hard to model \citep[e.g.][]{simmerer2003,bonifacio2010}.

\subsubsection{Neutron-capture elements: Sr, Y, Ba, Eu}

The neutron-capture elements are important for furthering our understanding of the formation of heavy elements, e.g. understanding the slow (s-) and rapid (r-) neutron capture processes, putting constraints on yields from AGB stars, SN, and neutron-star mergers, as well as improving our understanding of the overall chemical evolution of our Galaxy \citep[cf.][and references therein]{karakas2014}.  To make progress information on more than one heavy element is needed. It is important to study the distribution and behavior around both the first as well as the second rapid and slow neutron-capture process peaks.

Sr and Y are excellent tracers for the weak s-process \citep[e.g.][]{heil2009,pignatari2010}, while  Ba primarily traces the main s-process at solar metallicity \citep[e.g.][]{busso1999,kappeler2011}.  However, at low metallicity, these elements can be produced via the main r-process.  The ratio [Sr/Ba] is especially important since a large spread in this ratio has been observed in metal-poor stars in the halo, which has yet to be fully explained \citep[e.g.][]{francois2007}.

The lines for Sr and Y are quite weak, and Y lines are typically very blended (especially for $R<20,000$).  Ba, on the other hand, includes several strong lines.  Several lines, however, are affected by isotopic splitting, which can be difficult to constrain \citep[e.g.][]{mashonkina2006,gallagher2012}.  Eu is produced via the r-process, but only has a single good (and weak) line in the optical, at 6645~\AA.  The remaining heavy elements have lines that are too weak to measure.

\subsection{An aside: non-local thermodynamic equilibrium effects}

As with Fe, the lines of many elements can suffer from significant 3D and NLTE effects.  NLTE effects in FGK stars have already been studied in detail for many elements \citep[see, e.g.][and references therein]{mashonkina2014}.  It is beyond the scope and purpose of this paper to list the details of these effects, and how they can affect the interpretation of trends in Galactic chemical evolution models, for each individual element.  However, it is important to keep these issues in mind when developing the line list.

\subsection{The final line list}
\label{sec-finallist}

To source lines we used the following material as a starting point: working material from \citet{caffau2013}, \citet{bensby2014}, \citet{fulbright2000}, \citet{johnson2002}, and two line lists for short regions of horizontal branch stars in metal-rich globular clusters \citep[][Simmerer, priv. comm.]{feltzing2009}.  All of these lists include well-behaved lines used for successful analysis in the quoted studies (at their given $R$ and SNR).  In addition, we also included lines from the {\it Gaia}-ESO Survey (Heiter et al.~2014, in prep.) and the line list used for the optimization for HERMES (Freeman, priv. comm.).  A first list was compiled by simply merging these different line lists. The lines in this list were then critically evaluated, very weak lines were culled and each line was given a weight. 

The lines were first visually inspected in synthetic spectra convolved to a resolution of $R\sim20,000$ (ie. lower than in some of the studies used to source the list) for a turn-off star ($\teff\sim6500$~K), a main sequence star ($\teff\sim5800$~K), and a red clump star ($\teff\sim5000$~K), each covering three different metallicities (solar, $-0.5$~dex, $-1.0$~dex, respectively).  The lines were appraised according to three main criteria (strength, general appearance of the wavelength region and blends) and then given a weight based on this information. The strength was judged on how deep the decrement was at solar metallicity. In general, lines with a decrement of $0.8$ (where the continuum is at $1.0$) were kept. Potential blends were inspected and in particular we looked for a broadening or skewness of the line.  If a line was both deemed suitable for all three metallicities and types of stars then the weight was set to $w=1$. If only suitable for one stellar type, the weight was set to $w=2$. If the region is complex, due e.g. to telluric contamination or auto-ionizing bands, this was flagged.  Thereafter the lines were synthesized with and without blends at the same resolution and compared to very high-quality spectra of the Sun and Arcturus, to thoroughly assess their blending properties and astrophysical performance.  The weights of the lines were adjusted during this process. 

The final line list comprises all lines that were appraised between 4500~\AA{} and about 8000~\AA, and is shown in Table~\ref{tab-llist}.  Note that this line list does not include all lines that can give some abundance information. It includes only those lines that have been used in previous well-established studies and passed our inspection.  Obviously, this list can be augmented for specific purposes, for other types of objects and other science goals.  It can also be extended to bluer and redder wavelengths.

The first column gives the element species and ion, while the following three columns give the central wavelength ($\lambda$) in \AA, the excitation potential (EP), and oscillator strength (log(gf)), respectively.  The atomic data were compiled as part of the {\it Gaia}-ESO Survey (Heiter et al., in prep.), and the references for these values are given in the final column of the table.  

The final weight ($w$) prescribed to each line is listed in the fifth column, which is followed by several flags that contributed to this final weight.  The first flag ($f_{\rm gf}=1, 2, 3$) reflects the quality on the log(gf) value, where $f_{\rm gf}=1$ implies the highest quality.  The availability of Anstee, Barklem and O'Mara \citep[ABO;][]{anstee1991,anstee1995,barklem1997,barklem1998} cross-section data for collisional broadening by hydrogen for each line is given by the second flag, $f_{\rm b}=(0, 1)$.  A value of 1 equals `yes', while a value of 0 means `no'.  The last flag, $f_{\rm s}=(1, 2, 3)$, indicates the stellar type for which each line is measurable.  A value of 1 means the line can be measured in a dwarf-like star, while a value of 2 means the line can be measured in a giant.  If the line is measurable in both types of stars, $f_{\rm s}=3$. Further comments to some of the lines are given as footnotes to the table.

\centering
\footnotesize

 \begin{table*}
    \caption{Final compiled line list for the analysis of FGK stars in the Milky Way.}
 \begin{threeparttable}
 \begin{tabular}{ccccccccl}

 \hline\hline
 Species & $\lambda$ (\AA) & EP & log(gf) & $w$ & $f_{\rm gf}$ & $f_{\rm b}$ & $f_{\rm s}$ &   Ref., Comm.  \\
\hline
\nion{Mg}{I} & 5167.3216 & 2.709 & -0.931 & 2 & 1 & 1 & 3 & {\citet{ATJL}} \tnote{c}    \\
\nion{Mg}{I} & 5172.6843 & 2.712 & -0.450 & 1 & 1 & 1 & 3 & {\citet{ATJL}} \tnote{i}   \\
\nion{Mg}{I} & 5183.6042 & 2.717 & -0.239 & 1 & 1 & 1 & 3 & {\citet{ATJL}} \tnote{i}   \\
\nion{Ca}{I} & 6162.1730 & 1.899 & -0.090 & 1 & 1 & 1 & 3 & {\citet{SN}}     \\
\hline\hline
\nion{Li}{I} & 6707.7635 & 0.000 & -0.002 & 1 & 1 & 1 & 3 & {\citet{1998PhRvA..57.1652Y}}     \\
\nion{Li}{I} & 6707.9145 & 0.000 & -0.303 & 1 & 1 & 1 & 3 & {\citet{1998PhRvA..57.1652Y}}     \\
\nion{C}{I} & 5380.3252 & 7.685 & -1.615 & 1 & 1 & 1 & 1 & {\citet{1993A&AS...99..179H}}     \\
\nion{C}{I} & 6587.6100 & 8.537 & -1.021 & 1 & 1 & 1 & 1 & {\citet{1993A&AS...99..179H}}     \\
\nion{O}{I} & 6300.3038 & 0.000 & -9.715 & 2 & 1 & 0 & 2 & {\citet{2000MNRAS.312..813S,GESMCHF}} \tnote{a}    \\
\nion{O}{I} & 6363.7760 & 0.020 & -10.190 & 2 & 1 & 0 & 2 & {\citet{2000MNRAS.312..813S,GESMCHF}} \tnote{b}    \\
\nion{O}{I} & 7771.9440 & 9.146 & 0.369 & 2 & 2 & 1 & 3 & {\citet{1991JPhB...24.3943H}}     \\
\nion{O}{I} & 7774.1660 & 9.146 & 0.223 & 2 & 2 & 1 & 3 & {\citet{1991JPhB...24.3943H}}     \\
\nion{O}{I} & 7775.3880 & 9.146 & 0.002 & 2 & 2 & 1 & 3 & {\citet{1991JPhB...24.3943H}}     \\
\nion{Na}{I} & 5682.6333 & 2.102 & -0.706 & 2 & 2 & 1 & 3 & {\citet{GESMCHF}}     \\
\nion{Na}{I} & 5688.2050 & 2.104 & -0.404 & 1 & 2 & 1 & 3 & {\citet{GESMCHF}}     \\
\nion{Na}{I} & 5889.9509 & 0.000 & 0.108 & 1 & 1 & 1 & 3 & {\citet{1996PhRvL..76.2862V}}     \\
\nion{Na}{I} & 5895.9242 & 0.000 & -0.144 & 1 & 1 & 1 & 3 & {\citet{1996PhRvL..76.2862V}}     \\
\nion{Na}{I} & 6154.2255 & 2.102 & -1.547 & 1 & 2 & 0 & 3 & {\citet{GESMCHF}}     \\
\nion{Na}{I} & 6160.7471 & 2.104 & -1.246 & 1 & 2 & 0 & 3 & {\citet{GESMCHF}}     \\
\nion{Mg}{I} & 4571.0956 & 0.000 & -5.623 & 1 & 2 & 0 & 3 & {\citet{GESMCHF}}     \\
\nion{Mg}{I} & 4702.9909 & 4.346 & -0.440 & 1 & 2 & 1 & 3 & {\citet{1990JQSRT..43..207C}}     \\
\nion{Mg}{I} & 5528.4047 & 4.346 & -0.498 & 1 & 2 & 1 & 3 & {\citet{1990JQSRT..43..207C}}     \\
\nion{Mg}{I} & 5711.0880 & 4.346 & -1.724 & 1 & 2 & 0 & 3 & {\citet{1990JQSRT..43..207C}}     \\
\nion{Mg}{I} & 6318.7170 & 5.108 & -2.103 & 2 & 2 & 0 & 3 & {\citet{1993JPhB...26.4409B}}     \\
\nion{Mg}{I} & 6319.2370 & 5.108 & -2.324 & 2 & 2 & 0 & 3 & {\citet{1993JPhB...26.4409B}}     \\
\nion{Mg}{I} & 6319.4930 & 5.108 & -2.803 & 2 & 2 & 0 & 3 & {\citet{1993JPhB...26.4409B}}     \\
\nion{Mg}{I} & 7387.6890 & 5.753 & -1.000 & 1 & 2 & 0 & 3 & {\citet{1990JQSRT..43..207C}}     \\
\nion{Mg}{I} & 7691.5500 & 5.753 & -0.783 & 1 & 2 & 0 & 3 & {\citet{1990JQSRT..43..207C}}     \\
\nion{Al}{I} & 6698.6730 & 3.143 & -1.870 & 1 & 2 & 0 & 1 & {\citet{GESG12}}     \\
\nion{Al}{I} & 7835.3090 & 4.022 & -0.649 & 1 & 3 & 0 & 3 & {\citet{K07}}     \\
\nion{Al}{I} & 7836.1340 & 4.022 & -0.494 & 1 & 3 & 0 & 3 & {\citet{K07}}     \\
\nion{Si}{I} & 5665.5545 & 4.920 & -1.940 & 2 & 1 & 0 & 3 & {\citet{GARZ,BL}} \tnote{c}    \\
\nion{Si}{I} & 5690.4250 & 4.930 & -1.773 & 1 & 1 & 1 & 3 & {\citet{GARZ,BL}}     \\
\nion{Si}{I} & 5793.0726 & 4.930 & -1.963 & 1 & 1 & 1 & 3 & {\citet{GARZ,BL}}     \\
\nion{Si}{I} & 5948.5410 & 5.082 & -1.130 & 1 & 1 & 0 & 3 & {\citet{GARZ,BL}}     \\
\nion{Si}{I} & 6125.0209 & 5.614 & -1.464 & 2 & 3 & 0 & 1 & {\citet{K07}}     \\
\nion{Si}{I} & 6131.5729 & 5.616 & -1.556 & 1 & 3 & 0 & 3 & {\citet{K07}}     \\
\nion{Si}{I} & 6131.8516 & 5.616 & -1.615 & 2 & 3 & 0 & 3 & {\citet{K07}} \tnote{d} \\
\nion{Si}{I} & 6142.4832 & 5.619 & -1.295 & 2 & 3 & 0 & 3 & {\citet{K07}} \tnote{b}    \\
\nion{Si}{I} & 6145.0159 & 5.616 & -1.310 & 2 & 3 & 0 & 3 & {\citet{K07}} \tnote{e}   \\
\nion{Si}{I} & 6155.1343 & 5.619 & -0.754 & 1 & 3 & 0 & 3 & {\citet{K07}}     \\
\nion{Si}{I} & 6237.3191 & 5.614 & -0.975 & 1 & 3 & 0 & 3 & {\citet{K07}}     \\
\nion{Si}{I} & 6243.8146 & 5.616 & -1.242 & 2 & 3 & 0 & 1 & {\citet{K07}}     \\
\nion{Si}{I} & 6244.4655 & 5.616 & -1.093 & 2 & 3 & 0 & 1 & {\citet{K07}}     \\
\nion{Si}{I} & 6721.8481 & 5.863 & -1.062 & 2 & 2 & 0 & 3 & {\citet{1993PhyS...48..297N}}     \\
\nion{Si}{I} & 7034.9006 & 5.871 & -0.880 & 2 & 1 & 0 & 3 & {\citet{GARZ}}     \\
\nion{Si}{I} & 7250.6269 & 5.619 & -1.220 & 2 & 1 & 0 & 3 & {\citet{GARZ}} \tnote{e}   \\
\nion{Si}{I} & 7405.7718 & 5.614 & -0.820 & 1 & 1 & 0 & 3 & {\citet{GARZ}} \tnote{g}  \\
\nion{Si}{I} & 7423.4964 & 5.619 & -0.176 & 2 & 3 & 0 & 1 & {\citet{K07}}     \\
\nion{Si}{I} & 7680.2660 & 5.863 & -0.690 & 1 & 1 & 0 & 3 & {\citet{GARZ}}     \\
\nion{Si}{I} & 7849.9664 & 6.191 & -0.714 & 2 & 3 & 0 & 1 & {\citet{K07}}     \\
\nion{Si}{I} & 7918.3840 & 5.954 & -0.610 & 2 & 1 & 0 & 1 & {\citet{GARZ}}     \\
\nion{Si}{I} & 7932.3479 & 5.964 & -0.470 & 1 & 1 & 0 & 3 & {\citet{GARZ}}     \\
\nion{Si}{I} & 7944.0006 & 5.984 & -0.310 & 2 & 1 & 0 & 1 & {\citet{GARZ}}     \\
\nion{Si}{II} & 6347.1087 & 8.121 & 0.169 & 2 & 1 & 0 & 1 & {\citet{GARZ,BL}}     \\
\nion{Si}{II} & 6371.3714 & 8.121 & -0.044 & 2 & 1 & 0 & 1 & {\citet{GARZ,BL}}     \\
\hline\hline
\end{tabular}
\end{threeparttable}
\label{tab-llist}
\end{table*}

\centering
\footnotesize
 \begin{table*}
    \contcaption{Final compiled line list for the analysis of FGK stars in the Milky Way.}
 \begin{threeparttable}
 \begin{tabular}{ccccccccl}
 \hline\hline
 Species & $\lambda$ (\AA) & EP & log(gf) & $w$ & $f_{\rm gf}$ & $f_{\rm b}$ & $f_{\rm s}$ &   Ref., Comm.  \\
\hline
\nion{K}{I} & 7698.9643 & 0.000 & -0.154 & 1 & 3 & 1 & 3 & {\citet{K12}}     \\
\nion{Ca}{I} & 4578.5510 & 2.520 & -0.697 & 2 & 1 & 0 & 1 & {\citet{SR,Sm}}     \\
\nion{Ca}{I} & 5261.7040 & 2.521 & -0.579 & 2 & 1 & 1 & 3 & {\citet{SR}}     \\
\nion{Ca}{I} & 5512.9800 & 2.933 & -0.464 & 2 & 1 & 0 & 3 & {\citet{S}}     \\
\nion{Ca}{I} & 5581.9650 & 2.523 & -0.555 & 1 & 1 & 1 & 3 & {\citet{SR}} \tnote{g}  \\
\nion{Ca}{I} & 5590.1140 & 2.521 & -0.571 & 2 & 1 & 1 & 3 & {\citet{SR}}     \\
\nion{Ca}{I} & 5601.2770 & 2.526 & -0.523 & 1 & 1 & 1 & 3 & {\citet{SR}} \tnote{h}   \\
\nion{Ca}{I} & 5867.5620 & 2.933 & -1.570 & 1 & 1 & 0 & 3 & {\citet{S}} \tnote{b}    \\
\nion{Ca}{I} & 6122.2170 & 1.886 & -0.316 & 1 & 1 & 1 & 3 & {\citet{SN}}     \\
\nion{Ca}{I} & 6161.2970 & 2.523 & -1.266 & 2 & 1 & 1 & 3 & {\citet{SR}}     \\
\nion{Ca}{I} & 6166.4390 & 2.521 & -1.142 & 1 & 1 & 1 & 3 & {\citet{SR}}     \\
\nion{Ca}{I} & 6169.0420 & 2.523 & -0.797 & 1 & 1 & 1 & 3 & {\citet{SR}}     \\
\nion{Ca}{I} & 6169.5630 & 2.526 & -0.478 & 1 & 1 & 1 & 3 & {\citet{SR}}     \\
\nion{Ca}{I} & 6439.0750 & 2.526 & 0.390 & 1 & 1 & 1 & 3 & {\citet{SR}}     \\
\nion{Ca}{I} & 6455.5980 & 2.523 & -1.290 & 1 & 1 & 1 & 3 & {\citet{SR}}     \\
\nion{Ca}{I} & 6471.6620 & 2.526 & -0.686 & 1 & 1 & 1 & 3 & {\citet{SR}}     \\
\nion{Ca}{I} & 6493.7810 & 2.521 & -0.109 & 1 & 1 & 1 & 3 & {\citet{SR}}     \\
\nion{Ca}{I} & 6499.6500 & 2.523 & -0.818 & 1 & 1 & 1 & 2 & {\citet{SR}}     \\
\nion{Sc}{I} & 4743.8300 & 1.448 & 0.422 & 2 & 1 & 1 & 2 & {\citet{LD}} \tnote{g}  \\
\nion{Sc}{I} & 5484.6260 & 1.851 & 0.148 & 2 & 1 & 1 & 2 & {\citet{LD}} \tnote{g}  \\
\nion{Sc}{I} & 5520.4970 & 1.865 & 0.293 & 2 & 1 & 0 & 2 & {\citet{LD}} \tnote{g}  \\
\nion{Sc}{I} & 5671.8210 & 1.448 & 0.495 & 1 & 1 & 0 & 2 & {\citet{LD}}     \\
\nion{Sc}{I} & 6210.6580 & 0.000 & -1.529 & 1 & 1 & 1 & 2 & {\citet{LD}}     \\
\nion{Sc}{II} & 5239.8130 & 1.455 & -0.765 & 2 & 1 & 0 & 1 & {\citet{LD}}     \\
\nion{Sc}{II} & 5526.7900 & 1.768 & 0.024 & 1 & 1 & 0 & 3 & {\citet{LD}}     \\
\nion{Sc}{II} & 5657.8960 & 1.507 & -0.603 & 2 & 1 & 0 & 3 & {\citet{LD}}     \\
\nion{Sc}{II} & 5684.2020 & 1.507 & -1.074 & 2 & 1 & 0 & 3 & {\citet{LD}}     \\
\nion{Sc}{II} & 6279.7530 & 1.500 & -1.252 & 2 & 3 & 0 & 3 & {\citet{K09}}     \\
\nion{Sc}{II} & 6245.6366 & 1.507 & -1.022 & 2 & 3 & 0 & 3 & {\citet{K09}}     \\
\nion{Sc}{II} & 6604.6010 & 1.357 & -1.309 & 1 & 1 & 0 & 3 & {\citet{LD}}     \\
\nion{Ti}{I} & 4512.7344 & 0.836 & -0.400 & 2 & 1 & 1 & 3 & {\citet{2013ApJS..205...11L}}     \\
\nion{Ti}{I} & 4518.0220 & 0.826 & -0.250 & 2 & 1 & 1 & 3 & {\citet{2013ApJS..205...11L}}     \\
\nion{Ti}{I} & 4555.4839 & 0.848 & -0.400 & 2 & 1 & 1 & 3 & {\citet{2013ApJS..205...11L}}     \\
\nion{Ti}{I} & 4617.2688 & 1.749 & 0.440 & 1 & 1 & 1 & 3 & {\citet{2013ApJS..205...11L}}     \\
\nion{Ti}{I} & 4623.0972 & 1.739 & 0.160 & 2 & 1 & 1 & 3 & {\citet{2013ApJS..205...11L}}     \\
\nion{Ti}{I} & 4758.1180 & 2.249 & 0.510 & 2 & 1 & 1 & 3 & {\citet{2013ApJS..205...11L}}     \\
\nion{Ti}{I} & 4759.2696 & 2.256 & 0.590 & 2 & 1 & 1 & 3 & {\citet{2013ApJS..205...11L}}     \\
\nion{Ti}{I} & 4820.4110 & 1.503 & -0.380 & 1 & 1 & 1 & 3 & {\citet{2013ApJS..205...11L}}     \\
\nion{Ti}{I} & 4913.6136 & 1.873 & 0.220 & 2 & 1 & 1 & 3 & {\citet{2013ApJS..205...11L}}     \\
\nion{Ti}{I} & 4981.7304 & 0.848 & 0.570 & 1 & 1 & 1 & 3 & {\citet{2013ApJS..205...11L}}     \\
\nion{Ti}{I} & 5016.1613 & 0.848 & -0.480 & 2 & 1 & 1 & 3 & {\citet{2013ApJS..205...11L}}     \\
\nion{Ti}{I} & 5210.3850 & 0.048 & -0.820 & 2 & 1 & 1 & 1 & {\citet{2013ApJS..205...11L}}     \\
\nion{Ti}{I} & 5219.7000 & 0.021 & -2.220 & 1 & 1 & 1 & 3 & {\citet{2013ApJS..205...11L}}     \\
\nion{Ti}{I} & 5223.6200 & 2.092 & -0.490 & 2 & 1 & 1 & 2 & {\citet{NWL}}     \\
\nion{Ti}{I} & 5300.0107 & 1.053 & -2.300 & 2 & 1 & 1 & 2 & {\citet{2013ApJS..205...11L}}     \\
\nion{Ti}{I} & 5426.2500 & 0.021 & -2.950 & 1 & 1 & 1 & 2 & {\citet{1982MNRAS.199...21B}}     \\
\nion{Ti}{I} & 5471.1926 & 1.443 & -1.420 & 1 & 3 & 1 & 2 & {\citet{2013ApJS..205...11L}}     \\
\nion{Ti}{I} & 5689.4600 & 2.297 & -0.360 & 1 & 1 & 1 & 2 & {\citet{NWL}}     \\
\nion{Ti}{I} & 5716.4500 & 2.297 & -0.720 & 1 & 1 & 1 & 2 & {\citet{NWL}}     \\
\nion{Ti}{I} & 5866.4512 & 1.067 & -0.790 & 1 & 1 & 1 & 3 & {\citet{2013ApJS..205...11L}} \tnote{g}  \\
\nion{Ti}{I} & 5903.3149 & 1.067 & -2.089 & 1 & 1 & 1 & 2 & {\citet{1983MNRAS.204..883B}}     \\
\nion{Ti}{I} & 5918.5351 & 1.067 & -1.640 & 1 & 1 & 1 & 2 & {\citet{2013ApJS..205...11L}}     \\
\nion{Ti}{I} & 5922.1092 & 1.046 & -1.380 & 1 & 1 & 1 & 3 & {\citet{2013ApJS..205...11L}} \tnote{g}  \\
\nion{Ti}{I} & 5937.8089 & 1.067 & -1.940 & 1 & 1 & 1 & 2 & {\citet{2013ApJS..205...11L}}     \\
\nion{Ti}{I} & 5953.1596 & 1.887 & -0.273 & 2 & 1 & 1 & 3 & {\citet{1986MNRAS.220..289B}}     \\
\nion{Ti}{I} & 6064.6262 & 1.046 & -1.888 & 1 & 1 & 1 & 2 & {\citet{1983MNRAS.204..883B}}     \\
\nion{Ti}{I} & 6091.1713 & 2.267 & -0.320 & 1 & 1 & 1 & 3 & {\citet{2013ApJS..205...11L}}     \\
\nion{Ti}{I} & 6126.2160 & 1.067 & -1.368 & 1 & 1 & 1 & 3 & {\citet{1983MNRAS.204..883B}}     \\
\nion{Ti}{I} & 6258.1015 & 1.443 & -0.390 & 2 & 1 & 1 & 1 & {\citet{2013ApJS..205...11L}}     \\
\nion{Ti}{I} & 6261.0975 & 1.430 & -0.530 & 2 & 1 & 1 & 1 & {\citet{2013ApJS..205...11L}} \tnote{b}  \\
\hline\hline
\end{tabular}
\end{threeparttable}
\label{tab-llist}
\end{table*}

\centering
\footnotesize
 \begin{table*}
    \contcaption{Final compiled line list for the analysis of FGK stars in the Milky Way.}
 \begin{threeparttable}
 \begin{tabular}{ccccccccl}
 \hline\hline
 Species & $\lambda$ (\AA) & EP & log(gf) & $w$ & $f_{\rm gf}$ & $f_{\rm b}$ & $f_{\rm s}$ &   Ref., Comm.  \\
\hline
\nion{Ti}{I} & 6312.2359 & 1.460 & -1.550 & 1 & 1 & 1 & 2 & {\citet{2013ApJS..205...11L}}     \\
\nion{Ti}{I} & 6336.0985 & 1.443 & -1.690 & 1 & 1 & 1 & 2 & {\citet{2013ApJS..205...11L}}     \\
\nion{Ti}{I} & 6497.6838 & 1.443 & -2.020 & 1 & 1 & 0 & 2 & {\citet{2013ApJS..205...11L}}     \\
\nion{Ti}{I} & 6554.2229 & 1.443 & -1.150 & 1 & 1 & 1 & 2 & {\citet{2013ApJS..205...11L}} \tnote{f}  \\
\nion{Ti}{I} & 6743.1221 & 0.900 & -1.611 & 1 & 3 & 1 & 2 & {\citet{K10}}     \\
\nion{Ti}{II} & 4568.3140 & 1.224 & -3.030 & 2 & 1 & 0 & 2 & {\citet{RHL}}     \\
\nion{Ti}{II} & 4589.9580 & 1.237 & -1.620 & 2 & 1 & 0 & 3 & {\citet{RHL}}     \\
\nion{Ti}{II} & 4708.6621 & 1.237 & -2.350 & 2 & 1 & 0 & 3 & {\citet{2013AAS...22134804W}}     \\
\nion{Ti}{II} & 4779.9850 & 2.048 & -1.248 & 2 & 3 & 0 & 3 & {\citet{K10}}     \\
\nion{Ti}{II} & 4865.6114 & 1.116 & -2.700 & 2 & 1 & 0 & 3 & {\citet{2013AAS...22134804W}} \tnote{f}  \\
\nion{Ti}{II} & 4911.1948 & 3.124 & -0.640 & 2 & 1 & 0 & 3 & {\citet{2013AAS...22134804W}}     \\
\nion{Ti}{II} & 5185.9018 & 1.893 & -1.410 & 1 & 1 & 0 & 3 & {\citet{2013AAS...22134804W}} \tnote{g}  \\
\nion{Ti}{II} & 5211.5304 & 2.590 & -1.410 & 2 & 1 & 0 & 1 & {\citet{2013AAS...22134804W}}     \\
\nion{Ti}{II} & 5336.7710 & 1.582 & -1.600 & 1 & 1 & 0 & 3 & {\citet{2013AAS...22134804W}}     \\
\nion{Ti}{II} & 5381.0150 & 1.566 & -1.970 & 2 & 1 & 0 & 1 & {\citet{2013AAS...22134804W}}     \\
\nion{Ti}{II} & 5418.7510 & 1.582 & -2.130 & 1 & 1 & 0 & 3 & {\citet{2013AAS...22134804W}}     \\
\nion{Ti}{II} & 6559.5637 & 2.048 & -2.175 & 2 & 3 & 0 & 2 & {\citet{K10}} \tnote{f}  \\
\nion{V}{I} & 4577.1741 & 0.000 & -1.048 & 1 & 1 & 1 & 3 & {\citet{1985A&A...153..109W}}     \\
\nion{V}{I} & 5604.9312 & 1.043 & -1.280 & 1 & 1 & 1 & 2 & {\citet{1985A&A...153..109W}}     \\
\nion{V}{I} & 5668.3608 & 1.081 & -1.030 & 1 & 1 & 1 & 2 & {\citet{1985A&A...153..109W}}     \\
\nion{V}{I} & 5670.8527 & 1.081 & -0.420 & 1 & 1 & 1 & 3 & {\citet{1985A&A...153..109W}}     \\
\nion{V}{I} & 5703.5750 & 1.051 & -0.211 & 1 & 1 & 1 & 3 & {\citet{1985A&A...153..109W}}     \\
\nion{V}{I} & 5727.0480 & 1.081 & -0.012 & 1 & 1 & 1 & 3 & {\citet{1985A&A...153..109W}}     \\
\nion{V}{I} & 5737.0589 & 1.064 & -0.740 & 1 & 1 & 1 & 3 & {\citet{1985A&A...153..109W}}     \\
\nion{V}{I} & 6039.7219 & 1.064 & -0.650 & 1 & 1 & 1 & 3 & {\citet{1985A&A...153..109W}}     \\
\nion{V}{I} & 6058.1390 & 1.043 & -1.374 & 1 & 1 & 1 & 2 & {\citet{1985A&A...153..109W}}     \\
\nion{V}{I} & 6081.4410 & 1.051 & -0.579 & 2 & 1 & 1 & 2 & {\citet{1985A&A...153..109W}}     \\
\nion{V}{I} & 6090.2139 & 1.081 & -0.062 & 1 & 1 & 1 & 3 & {\citet{1985A&A...153..109W}}     \\
\nion{V}{I} & 6111.6445 & 1.043 & -0.715 & 1 & 1 & 1 & 2 & {\citet{1985A&A...153..109W}}     \\
\nion{V}{I} & 6119.5233 & 1.064 & -0.320 & 2 & 1 & 1 & 3 & {\citet{1985A&A...153..109W}}     \\
\nion{V}{I} & 6135.3608 & 1.051 & -0.746 & 2 & 1 & 1 & 2 & {\citet{1985A&A...153..109W}}     \\
\nion{V}{I} & 6150.1565 & 0.301 & -1.290 & 2 & 3 & 1 & 2 & {\citet{K09}}     \\
\nion{V}{I} & 6251.8273 & 0.287 & -1.340 & 1 & 1 & 1 & 2 & {\citet{1985A&A...153..109W}}     \\
\nion{V}{I} & 6274.6488 & 0.267 & -1.670 & 1 & 1 & 1 & 2 & {\citet{1985A&A...153..109W}}     \\
\nion{V}{I} & 6285.1499 & 0.275 & -1.510 & 1 & 1 & 1 & 2 & {\citet{1985A&A...153..109W}}     \\
\nion{V}{I} & 6292.8251 & 0.287 & -1.470 & 1 & 1 & 1 & 2 & {\citet{1985A&A...153..109W}}     \\
\nion{V}{I} & 6531.4146 & 1.218 & -0.840 & 1 & 1 & 1 & 2 & {\citet{1985A&A...153..109W}}     \\
\nion{Cr}{I} & 4545.9530 & 0.941 & -1.370 & 1 & 1 & 1 & 3 & {\citet{SLS}} \tnote{g}  \\
\nion{Cr}{I} & 4708.0130 & 3.168 & 0.070 & 2 & 1 & 1 & 3 & {\citet{SLS}}     \\
\nion{Cr}{I} & 5296.6910 & 0.983 & -1.360 & 1 & 1 & 1 & 3 & {\citet{SLS}}     \\
\nion{Cr}{I} & 5300.7450 & 0.983 & -2.000 & 2 & 1 & 1 & 3 & {\citet{SLS}}     \\
\nion{Cr}{I} & 5348.3150 & 1.004 & -1.210 & 1 & 1 & 1 & 3 & {\citet{SLS}}     \\
\nion{Cr}{I} & 5719.8160 & 3.013 & -1.580 & 2 & 1 & 1 & 2 & {\citet{SLS}} \tnote{b}    \\
\nion{Cr}{I} & 5783.0635 & 3.323 & -0.375 & 1 & 3 & 1 & 3 & {\citet{K10}}     \\
\nion{Cr}{I} & 5783.8497 & 3.322 & -0.295 & 2 & 3 & 1 & 3 & {\citet{MFW}}     \\
\nion{Cr}{I} & 5787.9180 & 3.322 & -0.083 & 1 & 3 & 1 & 3 & {\citet{MFW}}     \\
\nion{Cr}{I} & 6330.0910 & 0.941 & -2.787 & 1 & 3 & 1 & 3 & {\citet{K10}}     \\
\nion{Cr}{I} & 6537.9212 & 1.004 & -3.718 & 2 & 3 & 0 & 2 & {\citet{K10}}     \\
\nion{Cr}{I} & 6630.0109 & 1.030 & -3.560 & 2 & 3 & 1 & 2 & {\citet{MFW}}     \\
\nion{Cr}{II} & 4588.1990 & 4.071 & -0.627 & 2 & 1 & 1 & 3 & {\citet{PGBH}}     \\
\nion{Cr}{II} & 5237.3285 & 4.073 & -1.144 & 2 & 3 & 1 & 1 & {\citet{K10}}     \\
\nion{Cr}{II} & 5305.8526 & 3.827 & -2.363 & 2 & 3 & 1 & 1 & {\citet{K10}}     \\
\nion{Cr}{II} & 5313.5628 & 4.074 & -1.526 & 2 & 3 & 1 & 1 & {\citet{K10}}     \\
\nion{Mn}{I} & 4754.0400 & 2.282 & -0.080 & 1 & 1 & 0 & 3 & {\citet{DLSSC}}     \\
\nion{Mn}{I} & 4761.5100 & 2.953 & -0.274 & 2 & 1 & 1 & 3 & {\citet{DLSSC}}     \\
\nion{Mn}{I} & 4766.4200 & 2.920 & 0.105 & 2 & 1 & 1 & 3 & {\citet{DLSSC}}     \\
\nion{Mn}{I} & 4783.4270 & 2.298 & 0.044 & 2 & 1 & 0 & 3 & {\citet{DLSSC}}     \\
\nion{Mn}{I} & 5377.6073 & 3.844 & -0.166 & 2 & 3 & 0 & 3 & {\citet{K07}}     \\
\nion{Mn}{I} & 5394.6698 & 0.000 & -3.503 & 1 & 1 & 1 & 3 & {\citet{1984MNRAS.208..147B}}     \\
\nion{Mn}{I} & 5399.4745 & 3.853 & -0.345 & 2 & 3 & 0 & 3 & {\citet{K07}}     \\
\hline\hline
\end{tabular}
\end{threeparttable}
\label{tab-llist}
\end{table*}

\centering
\footnotesize
 \begin{table*}
    \contcaption{Final compiled line list for the analysis of FGK stars in the Milky Way.}
 \begin{threeparttable}
 \begin{tabular}{ccccccccl}
 \hline\hline
 Species & $\lambda$ (\AA) & EP & log(gf) & $w$ & $f_{\rm gf}$ & $f_{\rm b}$ & $f_{\rm s}$ &   Ref., Comm.  \\
\hline
\nion{Mn}{I} & 5420.3508 & 2.143 & -1.462 & 1 & 1 & 1 & 3 & {\citet{1984MNRAS.208..147B}}     \\
\nion{Mn}{I} & 5432.5392 & 0.000 & -3.795 & 2 & 1 & 1 & 3 & {\citet{1984MNRAS.208..147B}}     \\
\nion{Mn}{I} & 6013.5100 & 3.072 & -0.354 & 1 & 1 & 0 & 3 & {\citet{DLSSC}}     \\
\nion{Mn}{I} & 6016.6700 & 3.073 & -0.180 & 1 & 1 & 0 & 3 & {\citet{DLSSC}} \tnote{g}  \\
\nion{Mn}{I} & 6021.8200 & 3.075 & -0.054 & 1 & 1 & 0 & 3 & {\citet{DLSSC}}     \\
\nion{Fe}{I} & 4547.8470 & 3.547 & -1.012 & 2 & 1 & 1 & 1 & {\citet{BWL}}     \\
\nion{Fe}{I} & 4602.0007 & 1.608 & -3.134 & 1 & 1 & 1 & 3 & {\citet{BWL}}     \\
\nion{Fe}{I} & 4630.1200 & 2.279 & -2.587 & 2 & 1 & 1 & 3 & {\citet{BWL}}     \\
\nion{Fe}{I} & 4678.8457 & 3.603 & -0.833 & 2 & 1 & 1 & 3 & {\citet{BWL}}     \\
\nion{Fe}{I} & 4704.9478 & 3.686 & -1.470 & 1 & 2 & 1 & 3 & {\citet{MRW}}     \\
\nion{Fe}{I} & 4741.5294 & 2.832 & -1.765 & 2 & 1 & 1 & 1 & {\citet{BWL}}     \\
\nion{Fe}{I} & 4745.7998 & 3.654 & -1.270 & 2 & 1 & 0 & 1 & {\citet{BWL}}     \\
\nion{Fe}{I} & 4779.4391 & 3.415 & -2.020 & 2 & 1 & 1 & 1 & {\citet{BWL}}     \\
\nion{Fe}{I} & 4787.8266 & 2.998 & -2.557 & 2 & 1 & 1 & 1 & {\citet{BKK,BWL}}     \\
\nion{Fe}{I} & 4788.7566 & 3.237 & -1.763 & 2 & 1 & 1 & 3 & {\citet{BWL}}     \\
\nion{Fe}{I} & 4802.8800 & 3.642 & -1.514 & 2 & 3 & 0 & 3 & {\citet{BWL,K07}} \tnote{d}  \\
\nion{Fe}{I} & 4882.1431 & 3.417 & -1.598 & 2 & 3 & 1 & 3 & {\citet{K07}}     \\
\nion{Fe}{I} & 4892.8589 & 4.218 & -1.290 & 2 & 2 & 1 & 2 & {\citet{1970A&A.....9...37R,FMW}}     \\
\nion{Fe}{I} & 4903.3099 & 2.882 & -0.903 & 1 & 1 & 1 & 3 & {\citet{BWL,GESRHL14}} \tnote{k} \\
\nion{Fe}{I} & 4917.2299 & 4.191 & -1.080 & 1 & 2 & 1 & 3 & {\citet{MRW}}     \\
\nion{Fe}{I} & 4946.3880 & 3.368 & -1.110 & 2 & 1 & 1 & 3 & {\citet{GESRHL14}}     \\
\nion{Fe}{I} & 4962.5716 & 4.178 & -1.182 & 2 & 1 & 1 & 2 & {\citet{BWL}}     \\
\nion{Fe}{I} & 4969.9173 & 4.218 & -0.710 & 2 & 2 & 1 & 3 & {\citet{1969A&A.....2..274G,FMW}}     \\
\nion{Fe}{I} & 4994.1295 & 0.915 & -3.058 & 2 & 1 & 1 & 3 & {\citet{BKK,GESB79b,BWL}}     \\
\nion{Fe}{I} & 5001.8633 & 3.882 & -0.010 & 1 & 1 & 1 & 3 & {\citet{GESRHL14}}     \\
\nion{Fe}{I} & 5044.2108 & 2.851 & -2.038 & 2 & 1 & 1 & 1 & {\citet{BK,BWL}}     \\
\nion{Fe}{I} & 5054.6425 & 3.640 & -1.921 & 2 & 1 & 1 & 1 & {\citet{BWL}}     \\
\nion{Fe}{I} & 5067.1495 & 4.220 & -0.970 & 2 & 2 & 1 & 1 & {\citet{1970A&A.....9...37R,FMW}}     \\
\nion{Fe}{I} & 5083.3382 & 0.958 & -2.939 & 2 & 1 & 1 & 3 & {\citet{BKK,GESB79b,BWL}}     \\
\nion{Fe}{I} & 5090.7731 & 4.256 & -0.440 & 2 & 2 & 1 & 1 & {\citet{BKor,2006JPCRD..35.1669F}}     \\
\nion{Fe}{I} & 5127.3592 & 0.915 & -3.306 & 2 & 1 & 1 & 1 & {\citet{GESB79b,BWL}}     \\
\nion{Fe}{I} & 5141.7389 & 2.424 & -1.978 & 1 & 1 & 1 & 3 & {\citet{BKK,BWL}}     \\
\nion{Fe}{I} & 5159.0576 & 4.283 & -0.820 & 2 & 2 & 1 & 1 & {\citet{WBW,FMW}}     \\
\nion{Fe}{I} & 5198.7108 & 2.223 & -2.135 & 2 & 1 & 1 & 1 & {\citet{GESB82c,BWL}}     \\
\nion{Fe}{I} & 5217.3893 & 3.211 & -1.100 & 1 & 1 & 1 & 3 & {\citet{BKK,BWL}}     \\
\nion{Fe}{I} & 5225.5260 & 0.110 & -4.789 & 2 & 1 & 1 & 1 & {\citet{BIPS,BWL}}     \\
\nion{Fe}{I} & 5232.9400 & 2.940 & -0.070 & 1 & 1 & 1 & 3 & {\citet{BK,BWL}}     \\
\nion{Fe}{I} & 5242.4907 & 3.634 & -0.967 & 1 & 1 & 1 & 3 & {\citet{BWL}}     \\
\nion{Fe}{I} & 5243.7763 & 4.256 & -1.050 & 2 & 1 & 1 & 3 & {\citet{MRW}}     \\
\nion{Fe}{I} & 5250.2090 & 0.121 & -4.933 & 1 & 1 & 1 & 3 & {\citet{BIPS,BWL}}     \\
\nion{Fe}{I} & 5250.6456 & 2.198 & -2.180 & 2 & 1 & 1 & 1 & {\citet{BWL}}     \\
\nion{Fe}{I} & 5253.4617 & 3.283 & -1.573 & 2 & 1 & 1 & 1 & {\citet{BK}}     \\
\nion{Fe}{I} & 5288.5247 & 3.695 & -1.490 & 2 & 1 & 1 & 1 & {\citet{BWL,GESRHL14}}     \\
\nion{Fe}{I} & 5293.9588 & 4.143 & -1.770 & 2 & 2 & 1 & 3 & {\citet{MRW}} \tnote{g}  \\
\nion{Fe}{I} & 5295.3121 & 4.415 & -1.590 & 2 & 2 & 1 & 3 & {\citet{MRW}}     \\
\nion{Fe}{I} & 5302.3003 & 3.283 & -0.720 & 2 & 1 & 1 & 3 & {\citet{BKK}}     \\
\nion{Fe}{I} & 5322.0407 & 2.279 & -2.802 & 1 & 1 & 1 & 3 & {\citet{BWL}} \tnote{g}  \\
\nion{Fe}{I} & 5364.8709 & 4.446 & 0.228 & 1 & 1 & 1 & 3 & {\citet{BWL}}     \\
\nion{Fe}{I} & 5365.3987 & 3.573 & -1.020 & 1 & 1 & 1 & 3 & {\citet{BWL}}     \\
\nion{Fe}{I} & 5373.7086 & 4.473 & -0.710 & 1 & 1 & 1 & 3 & {\citet{GESRHL14}}     \\
\nion{Fe}{I} & 5379.5736 & 3.695 & -1.514 & 1 & 1 & 1 & 3 & {\citet{BWL}}     \\
\nion{Fe}{I} & 5383.3685 & 4.313 & 0.645 & 1 & 1 & 1 & 3 & {\citet{BWL}}     \\
\nion{Fe}{I} & 5386.3331 & 4.154 & -1.670 & 1 & 2 & 1 & 3 & {\citet{MRW}}     \\
\nion{Fe}{I} & 5389.4788 & 4.415 & -0.410 & 2 & 2 & 1 & 1 & {\citet{WBW,FMW}}     \\
\nion{Fe}{I} & 5398.2791 & 4.446 & -0.630 & 1 & 2 & 1 & 3 & {\citet{MRW}}     \\
\nion{Fe}{I} & 5415.1989 & 4.387 & 0.643 & 1 & 1 & 1 & 3 & {\citet{BWL}}     \\
\nion{Fe}{I} & 5417.0332 & 4.415 & -1.580 & 2 & 2 & 1 & 1 & {\citet{MRW}}     \\
\nion{Fe}{I} & 5434.5235 & 1.011 & -2.121 & 1 & 1 & 1 & 3 & {\citet{BKK,GESB79b,BWL}}     \\
\nion{Fe}{I} & 5441.3387 & 4.313 & -1.630 & 1 & 2 & 1 & 3 & {\citet{MRW}}     \\
\nion{Fe}{I} & 5445.0420 & 4.387 & -0.020 & 1 & 2 & 1 & 3 & {\citet{1970ApJ...162.1037W,FMW}}     \\
\nion{Fe}{I} & 5461.5495 & 4.446 & -1.800 & 2 & 2 & 1 & 1 & {\citet{MRW}}     \\
\hline\hline
\end{tabular}
\end{threeparttable}
\label{tab-llist}
\end{table*}

\centering
\footnotesize
 \begin{table*}
    \contcaption{Final compiled line list for the analysis of FGK stars in the Milky Way.}
 \begin{threeparttable}
 \begin{tabular}{ccccccccl}
 \hline\hline
 Species & $\lambda$ (\AA) & EP & log(gf) & $w$ & $f_{\rm gf}$ & $f_{\rm b}$ & $f_{\rm s}$ &   Ref., Comm.  \\
\hline
\nion{Fe}{I} & 5466.3958 & 4.371 & -0.630 & 2 & 2 & 1 & 3 & {\citet{WBW,FMW}} \tnote{g}  \\
\nion{Fe}{I} & 5466.9880 & 3.573 & -2.233 & 2 & 1 & 0 & 3 & {\citet{BWL}} \tnote{g}  \\
\nion{Fe}{I} & 5473.9005 & 4.154 & -0.720 & 2 & 1 & 1 & 1 & {\citet{GESRHL14}}     \\
\nion{Fe}{I} & 5501.4649 & 0.958 & -3.046 & 1 & 1 & 1 & 3 & {\citet{BWL}}     \\
\nion{Fe}{I} & 5506.7787 & 0.990 & -2.795 & 2 & 1 & 1 & 3 & {\citet{GESB79b,BWL}}     \\
\nion{Fe}{I} & 5522.4461 & 4.209 & -1.450 & 1 & 2 & 1 & 3 & {\citet{MRW}}     \\
\nion{Fe}{I} & 5525.5436 & 4.231 & -1.084 & 2 & 1 & 1 & 3 & {\citet{BK}} \tnote{g}  \\
\nion{Fe}{I} & 5543.9356 & 4.218 & -1.040 & 1 & 2 & 1 & 3 & {\citet{MRW}}     \\
\nion{Fe}{I} & 5546.5058 & 4.371 & -1.210 & 2 & 2 & 1 & 3 & {\citet{MRW}}     \\
\nion{Fe}{I} & 5560.2115 & 4.435 & -1.090 & 2 & 2 & 1 & 3 & {\citet{MRW}}     \\
\nion{Fe}{I} & 5569.6180 & 3.417 & -0.486 & 1 & 1 & 1 & 3 & {\citet{BK}}     \\
\nion{Fe}{I} & 5576.0888 & 3.430 & -0.900 & 1 & 2 & 1 & 3 & {\citet{MRW}}     \\
\nion{Fe}{I} & 5586.7555 & 3.368 & -0.114 & 1 & 1 & 1 & 3 & {\citet{BKK,BWL}}     \\
\nion{Fe}{I} & 5618.6323 & 4.209 & -1.275 & 2 & 1 & 1 & 1 & {\citet{BWL}}     \\
\nion{Fe}{I} & 5633.9461 & 4.991 & -0.230 & 1 & 2 & 1 & 3 & {\citet{MRW}}     \\
\nion{Fe}{I} & 5638.2621 & 4.220 & -0.720 & 1 & 1 & 1 & 3 & {\citet{GESRHL14}}     \\
\nion{Fe}{I} & 5651.4689 & 4.473 & -1.900 & 2 & 2 & 1 & 3 & {\citet{MRW}}     \\
\nion{Fe}{I} & 5652.3176 & 4.260 & -1.850 & 2 & 2 & 1 & 3 & {\citet{MRW}}     \\
\nion{Fe}{I} & 5679.0229 & 4.652 & -0.820 & 1 & 2 & 1 & 3 & {\citet{MRW}}     \\
\nion{Fe}{I} & 5691.4970 & 4.301 & -1.420 & 2 & 2 & 1 & 3 & {\citet{MRW}}     \\
\nion{Fe}{I} & 5701.5442 & 2.559 & -2.193 & 1 & 1 & 1 & 3 & {\citet{BK,GESB82d,BWL}}     \\
\nion{Fe}{I} & 5705.4642 & 4.301 & -1.355 & 1 & 1 & 1 & 3 & {\citet{BK}}     \\
\nion{Fe}{I} & 5731.7618 & 4.256 & -1.200 & 2 & 2 & 1 & 3 & {\citet{MRW}}     \\
\nion{Fe}{I} & 5741.8477 & 4.256 & -1.672 & 2 & 1 & 1 & 3 & {\citet{BWL}}     \\
\nion{Fe}{I} & 5753.1223 & 4.260 & -0.688 & 2 & 1 & 1 & 3 & {\citet{BWL}}     \\
\nion{Fe}{I} & 5775.0805 & 4.220 & -1.297 & 1 & 1 & 1 & 3 & {\citet{BWL}}     \\
\nion{Fe}{I} & 5778.4530 & 2.588 & -3.430 & 1 & 1 & 1 & 3 & {\citet{BK}}     \\
\nion{Fe}{I} & 5793.9147 & 4.220 & -1.600 & 2 & 2 & 1 & 3 & {\citet{MRW}}     \\
\nion{Fe}{I} & 5814.8071 & 4.283 & -1.870 & 2 & 2 & 1 & 3 & {\citet{MRW}}     \\
\nion{Fe}{I} & 5852.2187 & 4.549 & -1.230 & 2 & 2 & 1 & 1 & {\citet{MRW}}     \\
\nion{Fe}{I} & 5855.0758 & 4.608 & -1.478 & 1 & 1 & 1 & 3 & {\citet{BK}}     \\
\nion{Fe}{I} & 5905.6712 & 4.652 & -0.690 & 1 & 2 & 1 & 3 & {\citet{MRW}}     \\
\nion{Fe}{I} & 5909.9724 & 3.211 & -2.587 & 1 & 1 & 1 & 3 & {\citet{BK}} \tnote{g}  \\
\nion{Fe}{I} & 5916.2473 & 2.453 & -2.994 & 2 & 1 & 1 & 3 & {\citet{GESB82c,BWL}}     \\
\nion{Fe}{I} & 5927.7887 & 4.652 & -0.990 & 1 & 2 & 1 & 3 & {\citet{MRW}} \tnote{g}  \\
\nion{Fe}{I} & 5930.1799 & 4.652 & -0.230 & 1 & 2 & 1 & 3 & {\citet{WBW,FMW}}     \\
\nion{Fe}{I} & 5934.6545 & 3.929 & -1.070 & 2 & 2 & 1 & 3 & {\citet{MRW}}     \\
\nion{Fe}{I} & 5956.6940 & 0.859 & -4.599 & 1 & 1 & 1 & 3 & {\citet{GESB86,BWL}}     \\
\nion{Fe}{I} & 5984.8150 & 4.733 & -0.196 & 2 & 3 & 0 & 1 & {\citet{K07}}     \\
\nion{Fe}{I} & 5987.0648 & 4.796 & -0.429 & 1 & 3 & 0 & 3 & {\citet{K07}}     \\
\nion{Fe}{I} & 6003.0111 & 3.882 & -1.100 & 2 & 1 & 1 & 3 & {\citet{GESRHL14}}     \\
\nion{Fe}{I} & 6024.0575 & 4.549 & -0.120 & 1 & 2 & 1 & 3 & {\citet{WBW,FMW}}     \\
\nion{Fe}{I} & 6027.0508 & 4.076 & -1.089 & 1 & 1 & 1 & 3 & {\citet{BWL}}     \\
\nion{Fe}{I} & 6056.0046 & 4.733 & -0.320 & 1 & 1 & 1 & 3 & {\citet{GESRHL14}}     \\
\nion{Fe}{I} & 6065.4820 & 2.609 & -1.529 & 1 & 1 & 1 & 3 & {\citet{GESB82d,BWL}}     \\
\nion{Fe}{I} & 6079.0077 & 4.652 & -1.020 & 2 & 2 & 1 & 3 & {\citet{MRW}}     \\
\nion{Fe}{I} & 6093.6429 & 4.608 & -1.400 & 2 & 2 & 1 & 3 & {\citet{MRW}}     \\
\nion{Fe}{I} & 6096.6641 & 3.984 & -1.830 & 2 & 2 & 1 & 3 & {\citet{MRW}}     \\
\nion{Fe}{I} & 6127.9062 & 4.143 & -1.399 & 2 & 1 & 1 & 3 & {\citet{BWL}}     \\
\nion{Fe}{I} & 6137.6913 & 2.588 & -1.402 & 2 & 1 & 1 & 3 & {\citet{GESB82d,BWL}}     \\
\nion{Fe}{I} & 6151.6173 & 2.176 & -3.295 & 1 & 1 & 1 & 3 & {\citet{BKK,GESB82c,BWL}}     \\
\nion{Fe}{I} & 6157.7279 & 4.076 & -1.160 & 1 & 2 & 0 & 3 & {\citet{MRW}}     \\
\nion{Fe}{I} & 6165.3598 & 4.143 & -1.473 & 1 & 1 & 1 & 3 & {\citet{BWL}}     \\
\nion{Fe}{I} & 6173.3343 & 2.223 & -2.880 & 1 & 1 & 1 & 3 & {\citet{GESB82c}}     \\
\nion{Fe}{I} & 6180.2026 & 2.728 & -2.591 & 2 & 1 & 1 & 3 & {\citet{BK,BWL}}     \\
\nion{Fe}{I} & 6187.9892 & 3.943 & -1.620 & 1 & 2 & 1 & 3 & {\citet{MRW}}     \\
\nion{Fe}{I} & 6200.3125 & 2.609 & -2.433 & 1 & 1 & 1 & 3 & {\citet{GESB82d,BWL}}     \\
\nion{Fe}{I} & 6213.4294 & 2.223 & -2.481 & 2 & 1 & 1 & 3 & {\citet{BWL}}     \\
\nion{Fe}{I} & 6219.2805 & 2.198 & -2.432 & 1 & 1 & 1 & 3 & {\citet{BKK,GESB82c,BWL}}     \\
\nion{Fe}{I} & 6226.7342 & 3.884 & -2.120 & 2 & 2 & 1 & 3 & {\citet{MRW}}     \\
\nion{Fe}{I} & 6229.2259 & 2.845 & -2.805 & 2 & 1 & 1 & 3 & {\citet{BKK}}     \\
\nion{Fe}{I} & 6232.6403 & 3.654 & -1.223 & 2 & 1 & 0 & 3 & {\citet{BK}}     \\
\nion{Fe}{I} & 6246.3180 & 3.603 & -0.779 & 1 & 1 & 1 & 3 & {\citet{BKK,BWL}}     \\
\hline\hline
\end{tabular}
\end{threeparttable}
\label{tab-llist}
\end{table*}

\centering
\footnotesize
 \begin{table*}
    \contcaption{Final compiled line list for the analysis of FGK stars in the Milky Way.}
 \begin{threeparttable}
 \begin{tabular}{ccccccccl}
 \hline\hline
 Species & $\lambda$ (\AA) & EP & log(gf) & $w$ & $f_{\rm gf}$ & $f_{\rm b}$ & $f_{\rm s}$ &   Ref., Comm.  \\
\hline
\nion{Fe}{I} & 6252.5549 & 2.404 & -1.699 & 1 & 1 & 1 & 3 & {\citet{GESB82c,BWL}}     \\
\nion{Fe}{I} & 6265.1323 & 2.176 & -2.550 & 2 & 1 & 1 & 3 & {\citet{GESB82c,BWL}}     \\
\nion{Fe}{I} & 6270.2234 & 2.858 & -2.470 & 1 & 1 & 1 & 3 & {\citet{BKK,BWL}}     \\
\nion{Fe}{I} & 6297.7926 & 2.223 & -2.737 & 1 & 1 & 1 & 3 & {\citet{BKK,GESB82c,BWL}}     \\
\nion{Fe}{I} & 6322.6850 & 2.588 & -2.430 & 1 & 1 & 1 & 3 & {\citet{GESB82d,BWL}}     \\
\nion{Fe}{I} & 6335.3299 & 2.198 & -2.177 & 1 & 1 & 1 & 3 & {\citet{BWL}}     \\
\nion{Fe}{I} & 6336.8234 & 3.686 & -0.856 & 1 & 1 & 1 & 3 & {\citet{BK}}     \\
\nion{Fe}{I} & 6380.7432 & 4.186 & -1.375 & 2 & 1 & 1 & 3 & {\citet{BWL}}     \\
\nion{Fe}{I} & 6393.6004 & 2.433 & -1.452 & 1 & 1 & 1 & 3 & {\citet{BKK,BWL}}     \\
\nion{Fe}{I} & 6411.6480 & 3.654 & -0.634 & 1 & 1 & 1 & 3 & {\citet{BKK,BWL}}     \\
\nion{Fe}{I} & 6419.9487 & 4.733 & -0.200 & 2 & 2 & 1 & 3 & {\citet{MRW}}     \\
\nion{Fe}{I} & 6430.8450 & 2.176 & -2.005 & 1 & 1 & 1 & 3 & {\citet{GESB82c,BWL}}     \\
\nion{Fe}{I} & 6475.6239 & 2.559 & -2.941 & 1 & 1 & 1 & 3 & {\citet{BWL}}     \\
\nion{Fe}{I} & 6481.8698 & 2.279 & -2.981 & 1 & 1 & 1 & 3 & {\citet{BKK,GESB82c,BWL}}     \\
\nion{Fe}{I} & 6494.9804 & 2.404 & -1.268 & 1 & 1 & 1 & 3 & {\citet{GESB82c,BWL}}     \\
\nion{Fe}{I} & 6546.2381 & 2.759 & -1.536 & 1 & 1 & 1 & 3 & {\citet{BWL}}     \\
\nion{Fe}{I} & 6574.2266 & 0.990 & -5.004 & 2 & 1 & 1 & 2 & {\citet{GESB86,BWL}} \tnote{f}  \\
\nion{Fe}{I} & 6592.9124 & 2.728 & -1.473 & 1 & 1 & 1 & 3 & {\citet{BWL}}     \\
\nion{Fe}{I} & 6593.8695 & 2.433 & -2.420 & 1 & 1 & 1 & 3 & {\citet{GESB82c,BWL}}     \\
\nion{Fe}{I} & 6627.5438 & 4.549 & -1.590 & 1 & 1 & 1 & 3 & {\citet{GESRHL14}} \tnote{b}    \\
\nion{Fe}{I} & 6677.9851 & 2.692 & -1.418 & 1 & 1 & 1 & 3 & {\citet{BWL}}     \\
\nion{Fe}{I} & 6705.1009 & 4.607 & -0.870 & 1 & 1 & 0 & 3 & {\citet{GESRHL14}}     \\
\nion{Fe}{I} & 6713.7425 & 4.796 & -1.500 & 2 & 2 & 1 & 3 & {\citet{MRW}} \tnote{b}    \\
\nion{Fe}{I} & 6715.3818 & 4.608 & -1.540 & 2 & 2 & 1 & 1 & {\citet{MRW}} \tnote{b}    \\
\nion{Fe}{I} & 6725.3558 & 4.103 & -2.100 & 2 & 1 & 1 & 3 & {\citet{GESRHL14}} \tnote{b}    \\
\nion{Fe}{I} & 6726.6663 & 4.607 & -1.133 & 1 & 3 & 0 & 3 & {\citet{K07}}     \\
\nion{Fe}{I} & 6733.1503 & 4.638 & -1.480 & 1 & 2 & 1 & 3 & {\citet{MRW}}     \\
\nion{Fe}{I} & 6750.1515 & 2.424 & -2.618 & 1 & 1 & 1 & 3 & {\citet{BKK,GESB82c,BWL}}     \\
\nion{Fe}{I} & 6752.7066 & 4.638 & -1.204 & 2 & 1 & 1 & 1 & {\citet{BK}}     \\
\nion{Fe}{I} & 6806.8429 & 2.728 & -2.130 & 2 & 2 & 1 & 3 & {\citet{MRW}} \tnote{k}   \\
\nion{Fe}{I} & 6810.2622 & 4.607 & -0.986 & 1 & 1 & 1 & 3 & {\citet{BWL}}     \\
\nion{Fe}{I} & 6828.5912 & 4.638 & -0.820 & 1 & 2 & 1 & 3 & {\citet{MRW}}     \\
\nion{Fe}{I} & 6839.8300 & 2.559 & -3.350 & 1 & 2 & 1 & 3 & {\citet{MRW}}     \\
\nion{Fe}{I} & 6842.6853 & 4.638 & -1.220 & 1 & 2 & 1 & 3 & {\citet{MRW}}     \\
\nion{Fe}{I} & 6843.6554 & 4.549 & -0.730 & 1 & 1 & 1 & 3 & {\citet{GESRHL14}}     \\
\nion{Fe}{I} & 6857.2493 & 4.076 & -2.050 & 1 & 2 & 0 & 3 & {\citet{MRW}} \tnote{b}    \\
\nion{Fe}{I} & 6858.1483 & 4.608 & -0.930 & 1 & 1 & 1 & 3 & {\citet{BWL}}     \\
\nion{Fe}{I} & 7127.5676 & 4.988 & -1.046 & 2 & 3 & 0 & 1 & {\citet{K07}}     \\
\nion{Fe}{I} & 7132.9863 & 4.076 & -1.628 & 2 & 1 & 0 & 1 & {\citet{BWL}}     \\
\nion{Fe}{I} & 7418.6668 & 4.143 & -1.376 & 2 & 1 & 0 & 3 & {\citet{BWL}}     \\
\nion{Fe}{I} & 7491.6474 & 4.301 & -0.899 & 2 & 3 & 0 & 1 & {\citet{K07}}     \\
\nion{Fe}{I} & 7495.0656 & 4.220 & -0.100 & 2 & 1 & 0 & 1 & {\citet{GESRHL14}}     \\
\nion{Fe}{I} & 7568.8987 & 4.283 & -0.773 & 1 & 3 & 0 & 3 & {\citet{K07}}     \\
\nion{Fe}{I} & 7583.7881 & 3.018 & -1.885 & 1 & 1 & 1 & 3 & {\citet{BWL}}     \\
\nion{Fe}{I} & 7710.3632 & 4.220 & -1.113 & 2 & 1 & 0 & 3 & {\citet{BWL}}     \\
\nion{Fe}{I} & 7745.5133 & 5.086 & -1.172 & 2 & 3 & 0 & 1 & {\citet{K07}}     \\
\nion{Fe}{I} & 7746.5954 & 5.064 & -1.282 & 2 & 3 & 0 & 1 & {\citet{K07}}     \\
\nion{Fe}{I} & 7748.2693 & 2.949 & -1.751 & 2 & 1 & 1 & 3 & {\citet{BWL}}     \\
\nion{Fe}{I} & 7751.1090 & 4.991 & -0.753 & 1 & 3 & 0 & 3 & {\citet{K07}}     \\
\nion{Fe}{I} & 7807.9090 & 4.991 & -0.541 & 1 & 3 & 0 & 3 & {\citet{K07}}     \\
\nion{Fe}{I} & 7941.0879 & 3.274 & -2.286 & 1 & 1 & 1 & 3 & {\citet{BKK}}     \\
\nion{Fe}{II} & 4508.2803 & 2.856 & -2.440 & 2 & 1 & 1 & 3 & {\citet{2009A&A...497..611M}}     \\
\nion{Fe}{II} & 4576.3400 & 2.844 & -2.950 & 2 & 1 & 1 & 1 & {\citet{2009A&A...497..611M}}     \\
\nion{Fe}{II} & 4620.5128 & 2.828 & -3.210 & 1 & 1 & 1 & 1 & {\citet{2009A&A...497..611M}}     \\
\nion{Fe}{II} & 4629.3390 & 2.807 & -2.340 & 2 & 1 & 1 & 1 & {\citet{2009A&A...497..611M}}     \\
\nion{Fe}{II} & 4923.9212 & 2.891 & -1.260 & 1 & 1 & 1 & 3 & {\citet{2009A&A...497..611M}}     \\
\nion{Fe}{II} & 4993.3502 & 2.807 & -3.684 & 2 & 2 & 1 & 1 & {\citet{RU}}     \\
\nion{Fe}{II} & 5197.5675 & 3.231 & -2.220 & 2 & 1 & 1 & 1 & {\citet{2009A&A...497..611M}}     \\
\nion{Fe}{II} & 5234.6226 & 3.221 & -2.180 & 1 & 1 & 1 & 3 & {\citet{2009A&A...497..611M}}     \\
\nion{Fe}{II} & 5316.6087 & 3.153 & -1.870 & 1 & 1 & 1 & 3 & {\citet{2009A&A...497..611M}} \tnote{d} \\
\nion{Fe}{II} & 5325.5523 & 3.221 & -3.160 & 2 & 1 & 1 & 1 & {\citet{2009A&A...497..611M}}     \\
\nion{Fe}{II} & 5414.0698 & 3.221 & -3.580 & 1 & 1 & 1 & 3 & {\citet{2009A&A...497..611M}}     \\
\nion{Fe}{II} & 5425.2485 & 3.199 & -3.220 & 1 & 1 & 1 & 3 & {\citet{2009A&A...497..611M}}     \\
\hline\hline
\end{tabular}
\end{threeparttable}
\label{tab-llist}
\end{table*}

\centering
\footnotesize
 \begin{table*}
    \contcaption{Final compiled line list for the analysis of FGK stars in the Milky Way.}
 \begin{threeparttable}
 \begin{tabular}{ccccccccl}
 \hline\hline
 Species & $\lambda$ (\AA) & EP & log(gf) & $w$ & $f_{\rm gf}$ & $f_{\rm b}$ & $f_{\rm s}$ &   Ref., Comm.  \\
\hline
\nion{Fe}{II} & 5534.8380 & 3.245 & -2.865 & 2 & 2 & 1 & 1 & {\citet{RU}}     \\
\nion{Fe}{II} & 5991.3709 & 3.153 & -3.647 & 2 & 2 & 1 & 1 & {\citet{RU}}     \\
\nion{Fe}{II} & 6084.1017 & 3.199 & -3.881 & 2 & 2 & 1 & 3 & {\citet{RU}}     \\
\nion{Fe}{II} & 6149.2459 & 3.889 & -2.841 & 1 & 2 & 1 & 3 & {\citet{RU}}     \\
\nion{Fe}{II} & 6238.3859 & 3.889 & -2.600 & 2 & 1 & 1 & 1 & {\citet{2009A&A...497..611M}}     \\
\nion{Fe}{II} & 6247.5569 & 3.892 & -2.435 & 1 & 2 & 1 & 3 & {\citet{RU}} \tnote{d} \\
\nion{Fe}{II} & 6369.4590 & 2.891 & -4.110 & 2 & 1 & 1 & 1 & {\citet{2009A&A...497..611M}}     \\
\nion{Fe}{II} & 6416.9190 & 3.892 & -2.877 & 2 & 2 & 1 & 3 & {\citet{RU}} \tnote{j}  \\
\nion{Fe}{II} & 6432.6800 & 2.891 & -3.570 & 1 & 1 & 1 & 3 & {\citet{2009A&A...497..611M}}     \\
\nion{Fe}{II} & 6456.3796 & 3.903 & -2.185 & 1 & 2 & 1 & 3 & {\citet{RU}}     \\
\nion{Fe}{II} & 6516.0766 & 2.891 & -3.310 & 1 & 1 & 1 & 3 & {\citet{2009A&A...497..611M}}     \\
\nion{Fe}{II} & 7222.3912 & 3.889 & -3.260 & 2 & 1 & 1 & 1 & {\citet{2009A&A...497..611M}}     \\
\nion{Fe}{II} & 7224.4778 & 3.889 & -3.200 & 2 & 1 & 1 & 1 & {\citet{2009A&A...497..611M}}     \\
\nion{Fe}{II} & 7711.7204 & 3.903 & -2.500 & 1 & 1 & 1 & 3 & {\citet{2009A&A...497..611M}}     \\
\nion{Co}{I} & 4588.7294 & 0.432 & -3.820 & 2 & 3 & 0 & 2 & {\citet{K08}}     \\
\nion{Co}{I} & 4813.4764 & 3.216 & 0.120 & 1 & 3 & 0 & 3 & {\citet{K08}}     \\
\nion{Co}{I} & 5301.0410 & 1.710 & -1.940 & 2 & 1 & 1 & 2 & {\citet{1999ApJS..122..557N}}     \\
\nion{Co}{I} & 5331.4532 & 1.785 & -1.990 & 1 & 1 & 1 & 3 & {\citet{1999ApJS..122..557N}}     \\
\nion{Co}{I} & 5342.7006 & 4.021 & 0.741 & 2 & 3 & 0 & 1 & {\citet{K08}}     \\
\nion{Co}{I} & 5352.0397 & 3.576 & 0.060 & 1 & 1 & 1 & 3 & {\citet{1982ApJ...260..395C}}     \\
\nion{Co}{I} & 5647.2338 & 2.280 & -1.560 & 1 & 1 & 1 & 3 & {\citet{1982ApJ...260..395C}}     \\
\nion{Co}{I} & 6116.9902 & 1.785 & -2.490 & 1 & 1 & 1 & 2 & {\citet{1982ApJ...260..395C}}     \\
\nion{Co}{I} & 6454.9943 & 3.632 & -0.250 & 2 & 1 & 1 & 2 & {\citet{1982ApJ...260..395C}}     \\
\nion{Co}{I} & 6771.0343 & 1.883 & -1.970 & 1 & 1 & 1 & 3 & {\citet{1982ApJ...260..395C}}     \\
\nion{Ni}{I} & 4831.1690 & 3.606 & -0.321 & 2 & 1 & 1 & 3 & {\citet{WLa}}     \\
\nion{Ni}{I} & 4904.4118 & 3.542 & -0.016 & 2 & 3 & 0 & 3 & {\citet{K08}}  \\
\nion{Ni}{I} & 4953.2000 & 3.740 & -0.580 & 2 & 1 & 1 & 1 & {\citet{WLa}}     \\
\nion{Ni}{I} & 5010.9381 & 3.635 & -0.677 & 2 & 3 & 1 & 3 & {\citet{K08}}     \\
\nion{Ni}{I} & 5035.3570 & 3.635 & 0.290 & 1 & 1 & 1 & 3 & {\citet{WLa}}     \\
\nion{Ni}{I} & 5082.3441 & 3.658 & -0.439 & 2 & 3 & 1 & 1 & {\citet{K08}}     \\
\nion{Ni}{I} & 5084.0957 & 3.679 & -0.084 & 2 & 3 & 1 & 3 & {\citet{K08}}     \\
\nion{Ni}{I} & 5084.0957 & 3.679 & -0.084 & 2 & 3 & 1 & 3 & {\citet{K08}}     \\
\nion{Ni}{I} & 5115.3922 & 3.834 & -0.015 & 2 & 3 & 1 & 3 & {\citet{K08}}     \\
\nion{Ni}{I} & 5578.7183 & 1.676 & -2.640 & 1 & 1 & 1 & 3 & {\citet{1985JQSRT..33..307D}}     \\
\nion{Ni}{I} & 5587.8578 & 1.935 & -2.140 & 2 & 1 & 1 & 3 & {\citet{1985JQSRT..33..307D}}     \\
\nion{Ni}{I} & 5593.7355 & 3.898 & -0.682 & 1 & 3 & 1 & 3 & {\citet{K08}}     \\
\nion{Ni}{I} & 5748.3507 & 1.676 & -3.242 & 2 & 3 & 1 & 3 & {\citet{K08}}     \\
\nion{Ni}{I} & 5805.2166 & 4.167 & -0.579 & 1 & 3 & 1 & 3 & {\citet{K08}}     \\
\nion{Ni}{I} & 5846.9935 & 1.676 & -3.210 & 1 & 1 & 1 & 3 & {\citet{1985JQSRT..33..307D}} \tnote{g}  \\
\nion{Ni}{I} & 5996.7301 & 4.236 & -1.037 & 2 & 3 & 1 & 3 & {\citet{K08}} \tnote{b}    \\
\nion{Ni}{I} & 6007.3098 & 1.676 & -3.740 & 1 & 3 & 1 & 3 & {\citet{K08}}     \\
\nion{Ni}{I} & 6086.2815 & 4.266 & -0.410 & 1 & 3 & 1 & 3 & {\citet{K08}}     \\
\nion{Ni}{I} & 6108.1158 & 1.676 & -2.440 & 1 & 1 & 1 & 3 & {\citet{1985JQSRT..33..307D}}     \\
\nion{Ni}{I} & 6111.0703 & 4.088 & -0.865 & 1 & 3 & 1 & 3 & {\citet{K08}}     \\
\nion{Ni}{I} & 6128.9731 & 1.676 & -3.320 & 2 & 1 & 1 & 3 & {\citet{1985JQSRT..33..307D}}     \\
\nion{Ni}{I} & 6175.3665 & 4.089 & -0.389 & 1 & 3 & 1 & 3 & {\citet{K08}}     \\
\nion{Ni}{I} & 6176.8070 & 4.088 & -0.260 & 1 & 1 & 1 & 3 & {\citet{WLa}}     \\
\nion{Ni}{I} & 6177.2415 & 1.826 & -4.018 & 2 & 3 & 0 & 3 & {\citet{K08}}     \\
\nion{Ni}{I} & 6186.7109 & 4.105 & -0.880 & 1 & 3 & 1 & 3 & {\citet{K08}}     \\
\nion{Ni}{I} & 6204.6000 & 4.088 & -1.100 & 1 & 1 & 1 & 3 & {\citet{WLa}} \tnote{g}  \\
\nion{Ni}{I} & 6223.9810 & 4.105 & -0.910 & 2 & 1 & 1 & 3 & {\citet{WLa}} \tnote{g}  \\
\nion{Ni}{I} & 6314.6585 & 1.935 & -1.770 & 1 & 2 & 1 & 3 & {\citet{LWST}}     \\
\nion{Ni}{I} & 6327.5985 & 1.676 & -3.150 & 1 & 2 & 1 & 3 & {\citet{LWST}}     \\
\nion{Ni}{I} & 6378.2470 & 4.154 & -0.830 & 1 & 1 & 1 & 3 & {\citet{WLa}}     \\
\nion{Ni}{I} & 6482.7983 & 1.935 & -2.630 & 1 & 2 & 1 & 3 & {\citet{LWST}}     \\
\nion{Ni}{I} & 6532.8730 & 1.935 & -3.357 & 2 & 3 & 1 & 3 & {\citet{K08}}     \\
\nion{Ni}{I} & 6586.3098 & 1.951 & -2.746 & 1 & 3 & 1 & 3 & {\citet{K08}}     \\
\nion{Ni}{I} & 6635.1224 & 4.419 & -0.765 & 2 & 3 & 1 & 1 & {\citet{K08}}     \\
\nion{Ni}{I} & 6643.6303 & 1.676 & -2.300 & 1 & 2 & 1 & 3 & {\citet{LWST}}     \\
\nion{Ni}{I} & 6767.7720 & 1.826 & -2.170 & 1 & 2 & 0 & 3 & {\citet{LWST}}     \\
\nion{Ni}{I} & 6772.3149 & 3.658 & -0.797 & 1 & 3 & 1 & 3 & {\citet{K08}}     \\
\nion{Ni}{I} & 6842.0367 & 3.658 & -1.374 & 2 & 3 & 1 & 1 & {\citet{K08}}     \\
\nion{Ni}{I} & 7110.8961 & 1.935 & -2.895 & 2 & 3 & 1 & 1 & {\citet{K08}}     \\
\hline\hline
\end{tabular}
\end{threeparttable}
\label{tab-llist}
\end{table*}

\centering
\footnotesize
 \begin{table*}
    \contcaption{Final compiled line list for the analysis of FGK stars in the Milky Way.}
 \begin{threeparttable}
 \begin{tabular}{ccccccccl}
 \hline\hline
 Species & $\lambda$ (\AA) & EP & log(gf) & $w$ & $f_{\rm gf}$ & $f_{\rm b}$ & $f_{\rm s}$ &   Ref., Comm.  \\
\hline
\nion{Ni}{I} & 7748.8843 & 3.706 & -0.185 & 1 & 3 & 0 & 3 & {\citet{K08}}     \\
\nion{Ni}{I} & 7788.9299 & 1.951 & -2.420 & 1 & 2 & 1 & 3 & {\citet{LWST}}     \\
\nion{Ni}{I} & 7797.5798 & 3.898 & -0.185 & 1 & 3 & 0 & 3 & {\citet{K08}}     \\
\nion{Cu}{I} & 5105.5370 & 1.389 & -1.516 & 2 & 1 & 0 & 1 & {\citet{KR,1989ZPhyD..11..287C}}     \\
\nion{Cu}{I} & 5700.2373 & 1.642 & -2.330 & 2 & 1 & 0 & 3 & {\citet{KR,1989ZPhyD..11..287C}}     \\
\nion{Cu}{I} & 5782.1269 & 1.642 & -1.781 & 1 & 1 & 0 & 3 & {\citet{KR,1989ZPhyD..11..287C}}     \\
\nion{Zn}{I} & 4722.1530 & 4.030 & -0.390 & 1 & 1 & 0 & 3 & {\citet{1980A&A....84..361B}}     \\
\nion{Zn}{I} & 4810.5280 & 4.078 & -0.160 & 1 & 1 & 1 & 1 & {\citet{1980A&A....84..361B}}     \\
\nion{Sr}{I} & 4607.3310 & 0.000 & 0.283 & 1 & 1 & 0 & 3 & {\citet{PRT}}     \\
\nion{Y}{II} & 4883.6821 & 1.084 & 0.265 & 1 & 3 & 0 & 3 & {\citet{K11}}     \\
\nion{Y}{II} & 5087.4160 & 1.084 & -0.170 & 1 & 1 & 0 & 1 & {\citet{HLGBW}}     \\
\nion{Y}{II} & 5200.4060 & 0.992 & -0.570 & 2 & 1 & 0 & 1 & {\citet{HLGBW}}     \\
\nion{Y}{II} & 5289.8150 & 1.033 & -1.850 & 2 & 1 & 0 & 2 & {\citet{HLGBW}}     \\
\nion{Y}{II} & 5662.9241 & 1.944 & 0.384 & 2 & 3 & 0 & 3 & {\citet{K11}}     \\
\nion{Ba}{II} & 4554.0290 & 0.000 & 0.140 & 1 & 1 & 1 & 3 & {\citet{1992A&A...255..457D}}     \\
\nion{Ba}{II} & 4934.0760 & 0.000 & -0.157 & 2 & 1 & 1 & 3 & {\citet{1992A&A...255..457D}}     \\
\nion{Ba}{II} & 5853.6680 & 0.604 & -0.907 & 1 & 1 & 1 & 3 & {\citet{1992A&A...255..457D}}     \\
\nion{Ba}{II} & 6141.7130 & 0.704 & -0.032 & 1 & 1 & 1 & 3 & {\citet{1992A&A...255..457D}}     \\
\nion{Ba}{II} & 6496.8970 & 0.604 & -0.407 & 1 & 1 & 1 & 3 & {\citet{1992A&A...255..457D}}     \\
\nion{Eu}{II} & 6645.0940 & 1.380 & 0.120 & 1 & 1 & 0 & 3 & {\citet{LWHS}}     \\
\hline\hline
\end{tabular}
\begin{tablenotes}
\item [a] Tellurics.
\item [b] Weak.
\item [c] Blended.
\item [d] Two blended lines of same species.
\item [e] Partly or weakly blended.
\item [f] Wing of a strong line.
\item [g] Overestimated theoretical blend.
\item [h] Unidentified blend.
\item [i] Continuum issue.
\item [j] Blended with \nion{Fe}{I}.
\item [k] Possible bad log(gf).
\item [l] Ca autoionization.
\end{tablenotes}
\end{threeparttable}
\label{tab-llist}
\end{table*}
%

\clearpage
\end{document}